\documentclass[12pt]{article}

\usepackage{authblk}	

\usepackage[margin=1in]{geometry} 

\usepackage[english]{babel} 
\usepackage{indentfirst}	

\usepackage[protrusion=true,expansion=true]{microtype} 
\usepackage{caption}

\usepackage{fancyhdr} 
\usepackage[numbers,round,sort&compress]{natbib}	

\newcommand{\beginsupplement}{
        \setcounter{table}{0}
        \renewcommand{\thetable}{S\arabic{table}}%
        \setcounter{figure}{0}
        \renewcommand{\thefigure}{S\arabic{figure}}%
     }

\usepackage{graphicx}
\usepackage{subcaption}	
\usepackage[section]{placeins}	
\usepackage{array,multirow}	
\usepackage{color,colortbl}	

\usepackage{amssymb,amsmath}

\usepackage{tocloft}

\cftsetindents{figure}{0em}{1in}
\cftpagenumbersoff{figure}
\cftsetpnumwidth{0em}
\cftsetrmarg{0em}
\title{DNA Base Pair Mismatches Induce Structural Changes and Alter the Free Energy Landscape of Base Flip}
\date{\today}
\author[1]{A. Kingsland}
\author[1]{L. Maibaum}
\affil[1]{Department of Chemistry, University of Washington, Seattle, WA, USA}

\begin{document}


\maketitle

\begin{center}
 \textit{\textbf{Running Title:}} DNA Mismatches and Structural Properties
 
 \medskip

\end{center}

\pagebreak


\section{Abstract}

Double-stranded DNA may contain mismatched base pairs beyond the Watson-Crick pairs guanine-cytosine and adenine-thymine.
Such mismatches bear adverse consequences for human health.
We utilize molecular dynamics and metadynamics computer simulations to study the equilibrium structure and dynamics for both matched and mismatched base pairs.
We discover significant differences between matched and mismatched pairs in structure, hydrogen bonding, and base flip work profiles.
Mismatched pairs shift further in the plane normal to the DNA strand and are more likely to exhibit non-canonical structures, including the e-motif.
We discuss potential implications on mismatch repair enzymes' detection of DNA mismatches.


\section{Introduction}

Bases in double-stranded DNA occur in complementary pairs: adenine (A) matches with thymine (T), and cytosine (C) matches with guanine (G).
This exact pairing allows for DNA duplication during mitosis and cell growth.
DNA replication, however, is susceptible to errors and mutations at a frequency of 1~per 10$^7$ base pairs per mitosis cycle;~\cite{hu_synergism_2017} this creates base pair mismatches, such as guanine-thymine (GT).
DNA mismatches induce negative consequences on human health, including higher mutation rates,~\cite{wind_hnpcc-like_1999} genetic defects,~\cite{kunkel_dna_2005,iyer_dna_2006,li_mechanisms_2008} and cancer.~\cite{li_mechanisms_2008,kolodner_human_1994,harfe_dna_2000,stivers_mechanistic_2003}

We study the effects of base pair mismatches on DNA structure at the molecular level.
We employ molecular dynamics computer simulations to quantify  the differences in structure between DNA strands with matched and mismatched base pairs.
Our focus is on geometric order parameters describing base pair structure,~\cite{yoon_structure_1988} hydrogen bonding between opposing bases, and flipping DNA bases out of the helix.
For a subset of these observables matched and mismatched base pairs deviate significantly, especially those which report on in-plane geometries.

To elucidate the physical properties of mismatched base pairs, we compute the work required to rotate a single base out of the DNA double helix.
This rotation, called base flip,~\cite{song_improved_2009}
is a crucial process by which enzymes detect and selectively repair post-replicative errors.
Proteins of the Mismatch Repair (MMR) system proofread the DNA; upon encountering a mismatch, an MMR enzyme flips the offending nucleotide out of the DNA double helix and into its active site before proceeding with the repair pathway.~\cite{lau_molecular_2000,harfe_dna_2000,stivers_extrahelical_2008,li_mechanisms_2008,law_base-flipping_2011}
MMR proteins decrease the mutation rate between 50-1000 times, to a rate of 1~per 10$^9$-10$^{10}$ base pairs per cell division.\cite{hu_synergism_2017,drake_distribution_1999}

Base flip requires an activation energy exceeding thermal fluctuations, and is unlikely to occur on the timescale of typical molecular dynamics simulations.
To overcome this obstacle, we utilize the well-tempered metadynamics algorithm; driving the simulation away from the ground state.
The free energy profiles in some mismatched pairs demonstrate a metastable state in which both bases flip out of the DNA strand and nest in the minor groove.
This may contribute to MMR proteins' ability to recognize mismatches.
Our results do not indicate a systematic trend in free energy difference between the flipped-in and flipped-out states as a quantitative predictor of mismatch recognition, largely due to this unusual minor-groove-nested structure.
The presence of this structure does not predict MMR recognition, which highlights the importance of analyzing additional physical properties such as DNA bending.~\cite{sharma_differential_2014,sharma_dna_2013,law_base-flipping_2011,erie_single_2014,vafabakhsh_extreme_2012}


\section{Methods}

\subsection{Model Building}

We construct atomistic models of B-DNA  containing twelve base pairs by employing the 3DNA software.~\cite{lu_3dna:_2008}
The sequence is of the form CTGA AC\underline{X}A ATGT, where the placeholder \underline{X} represents any one of the four nucleotides.
Bases 1-6 and 8-12 pair with matched nucleotides, whereas \underline{X} pairs with either a matched or a mismatched base, denoted Y. 
We use the notation XY to identify our considered sequence.
We study a total of eleven systems,
four matched ($\text{XY} \in \{\text{AT, TA, CG, GC}\}$) and seven with a single mismatch ($\text{XY} \in \{\text{AA, AC, AG, CC, CT, GT, TT}\}$).
We do not include the GG mismatch because steric influences prevent a hybridized equilibrium state.

We minimize each model for 200 steps of steepest descent by applying the CHARMM27 Force Field,~\cite{mackerell_development_2000} which we employed throughout.
We solvated the structures with explicit TIP3P water~\cite{jorgensen_comparison_1983} in a dodecahedral simulation box extending at least 1 nm beyond the DNA polymer in all directions.
We randomly substituted 22 water molecules for Na$^+$ counter-ions to neutralize the system.
The solvent was minimized for 200 steps of steepest descent and equilibrated for 100 ps.
We equilibrated each full system for 1 ns.

\subsection{Molecular Dynamics Simulations}

We performed simulations at 300 K in the NVT ensemble with stochastic velocity rescaling,~\cite{bussi_canonical_2007}
periodic boundary conditions, and a 2 fs timestep
employing the Gromacs simulation suite.~\cite{abraham_gromacs_2015}
We utilized the SHAKE algorithm~\cite{miyamoto_settle:_1992} to constrain covalent bonds that include a hydrogen atom,
and treated long-range electrostatic interactions with the Particle Mesh Ewald method~\cite{essmann_smooth_1995}
with a non-bonded cutoff of 1.2 nm.
After equilibration, we performed unbiased simulations for 100 ns and saved trajectory coordinates every 10 ps.
To identify representative configurations, we applied the Biopython~\cite{cock_biopython:_2009} module \texttt{cluster.kmedoids}.

\subsection{Structure, Hydrogen Bonding, and Base Flip}

\begin{figure}[htbp]
 \centering
 \includegraphics[width=\textwidth]{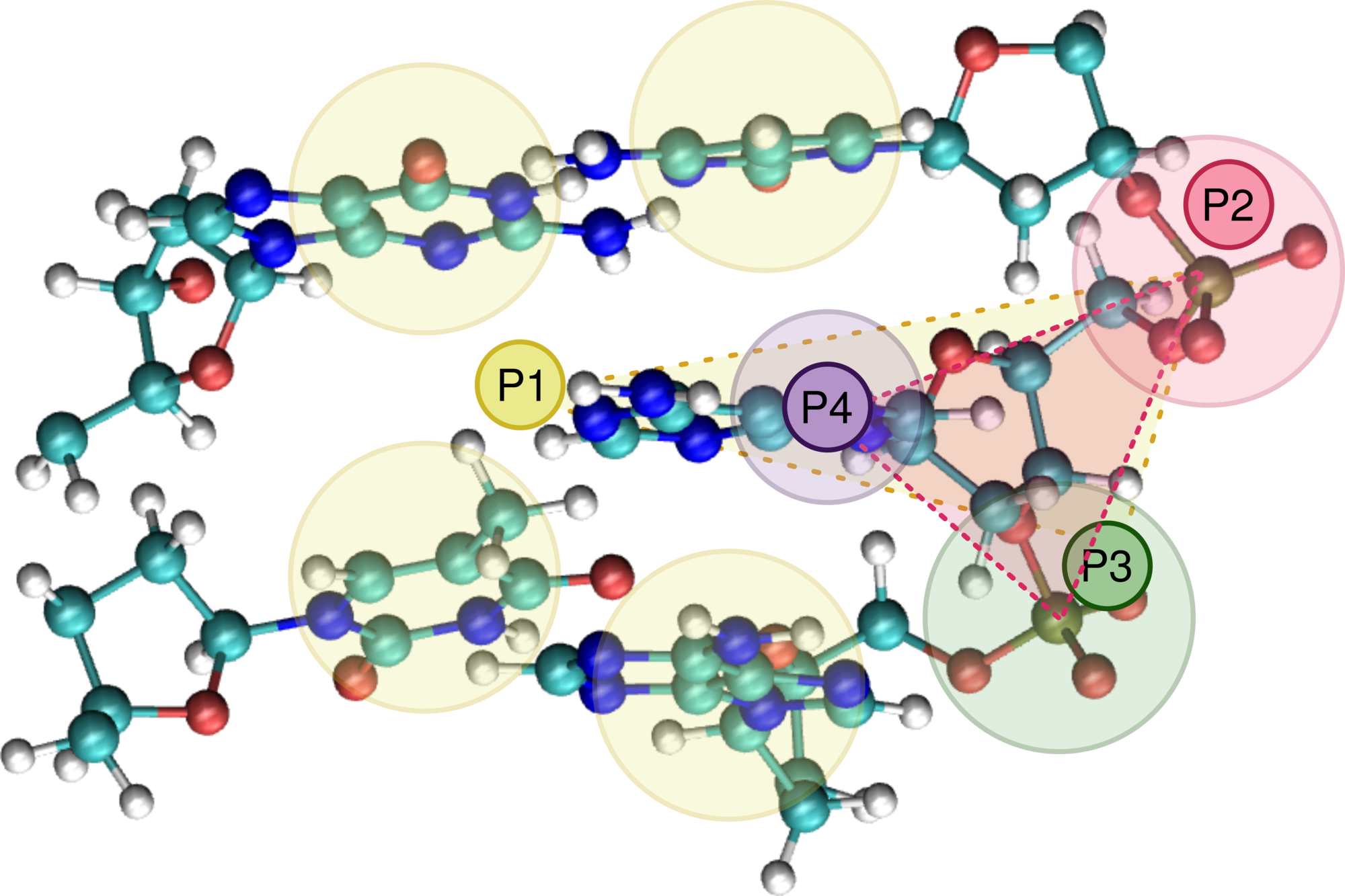}
 \caption{
 Definition of the base flip angle.
 We define the pseudo-dihedral angle as being between points P1, P2, P3, and P4;
 P1 is the center of mass of the neighboring base pairs (4 bases in total),
 P2 and P3 are the centers of mass of the phosphates flanking the flipping base,
 and P4 is the center of mass of the five-membered ring (of a purine) or the six-membered ring (of a pyrimidine).
 The two intersecting planes describing the dihedral angle are lightly shaded in yellow and red, and are outlined with dashed lines.
 For clarity, the opposing base is not pictured.}
 \label{fig:baseflip_dihedral}
\end{figure}

\begin{figure}[h]
\centering
 \includegraphics[width=0.75\textwidth]{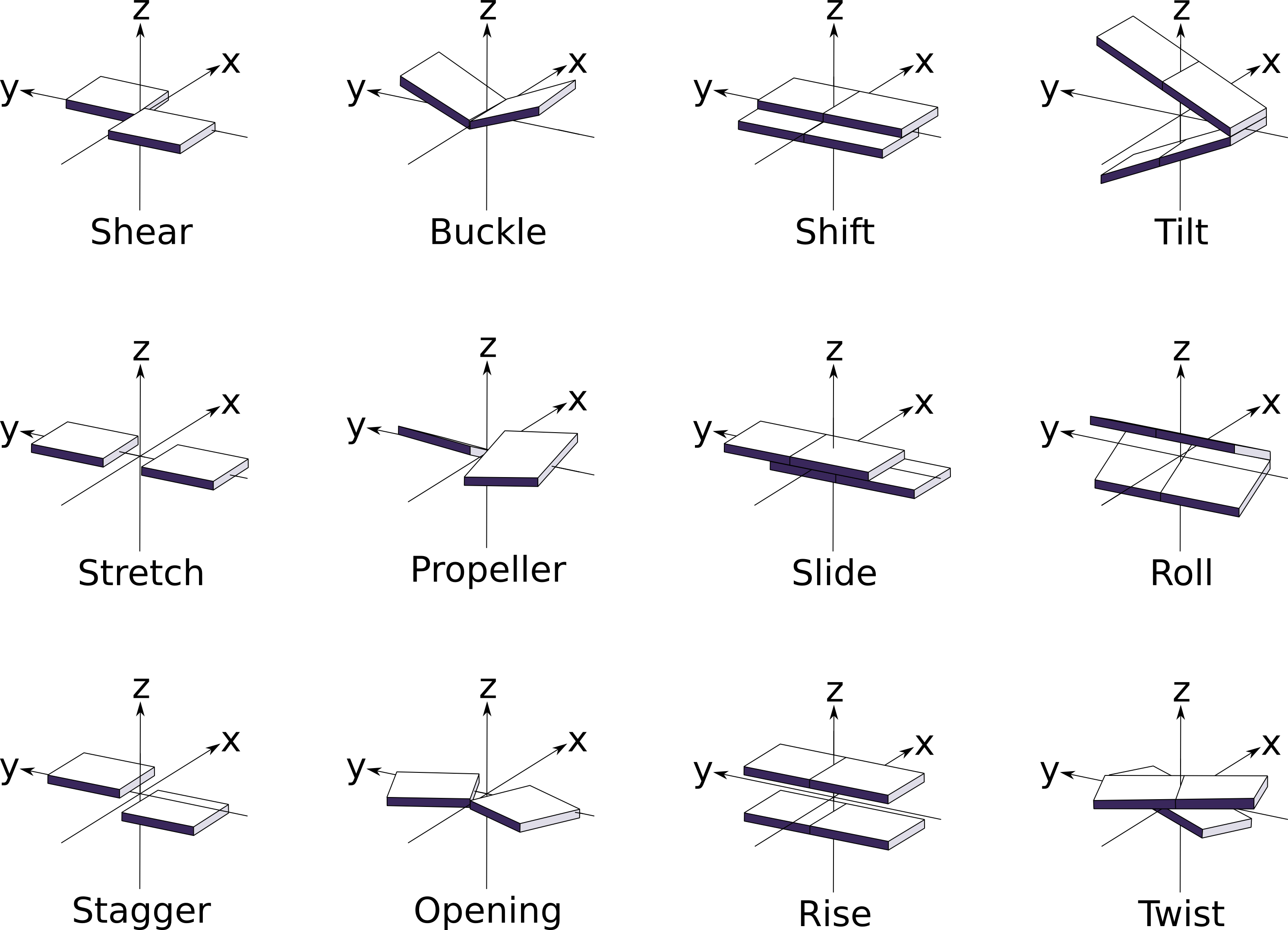}
 \caption{The structural parameters employed in this paper.  Adapted from Lu et. al. (1).
	  Parameters measured between a base and its pair are located in the left two columns, between a base pair and the neighboring pair are in the right two columns.
	  Translational parameters are in the first and third columns, rotational in the second and fourth.
}
 \label{fig:12orderparameters}
\end{figure}

Roll, tilt, twist, slide, rise, shift, buckle, propeller, opening, stagger, shear, stretch, (pictured in Fig.~\ref{fig:12orderparameters}) and inter-base hydrogen bonding were calculated using 3DNA;~\cite{lu_3dna:_2008}
probability distributions and free energy profiles were determined by in-house computer code.
We measure bend as presented by Sharma et. al.:~\cite{sharma_dna_2013} we section the DNA into three (bases 2-4/21-31, 5-8/17-20, and 9-11/14-16) and calculate the angle between the centers of mass of the heavy atoms.
To measure base flip, we adopt a pseudo-dihedral angle first described by Song et. al.,~\cite{song_improved_2009} defined as the dihedral angle between points P1, P2, P3, and P4
where P1 is the center of mass of the two neighboring base pairs, P2 and P3 are the centers of mass of the phosphates which flank the flipping base,
and P4 is the center of mass of the pyrimidine ring, or the five-membered ring of the purine.
(Fig.~\ref{fig:baseflip_dihedral})
Changes in base flip angle are positive if X crosses into the major groove or if Y crosses into the minor groove, and negative otherwise.

\subsection{Metadynamics Simulations}

We employ well-tempered metadynamics simulations with two collective variables: the base flip angle of X and the base flip angle of Y.
For comparison, we perform simulations in which we bias only one of these two base flip angles as a collective variable.
Using the PLUMED plug-in,~\cite{tribello_plumed_2014}
we add Gaussian biasing potentials to the collective variable(s) with a height of 3 kJ/mol, width of 0.08 radians, a deposition rate of 1 ps, and a well-tempered bias factor of 6, for a total simulation length of at least 700 ns per system.
We reconstruct the 2D free energy surface as the inverse of the potential energy added to the system as described in Ref.~\cite{barducci_well-tempered_2008}.
We pinned the base pairs terminating the DNA strand and the base pairs neighboring the flipping bases by their inter-pair hydrogen bonds to prevent the system from artificially melting.
We constrain each hydrogen bond to a distance of 0.2 nm with a harmonic constant of \mbox{40 kJ mol$^{-1}$ nm$^{-2}$}.

\subsection{Evaluating Correlations}

For any two order parameters $x$ and $y$, we calculated the covariance as
\begin{equation} \label{eq:normcov}
 \textrm{cov}_{xy} = \frac{\sigma_{xy}}{\sqrt{\sigma_{xx} \sigma_{yy}}}
\end{equation}
where
\begin{equation}
 \sigma_{xy} = \langle xy \rangle - \langle x \rangle \langle y \rangle .
\end{equation}
Here, the angular brackets denote the equilibrium average,
\begin{equation}
 \langle x \rangle = \iint_{-\pi}^{\pi} \textrm{d}x \textrm{d}y \, x P(x,y) ,
\end{equation}
where the probability distribution, $P$, is proportional to the exponent of the negative free energy:
\begin{equation}
P(x,y) \propto \exp \left(\frac{-G(x,y)}{k_\textrm{B} T}\right) .
\end{equation}
For well-tempered metadynamics simulations, we reconstructed free energy diagrams for structural order parameters utilizing the re-weighing method described by Bonomi et. al.~\cite{bonomi_reconstructing_2009}

\subsection{Calculating free energy differences}

We convert two-dimensional free energy profiles to one-dimensional free energy profiles by integrating:
\begin{equation} \label{eq:2Dto1D}
 G(x) = -k_\textrm{B} T \log \left[ \int_{-\infty}^{\infty} \textrm{d}y \, e^{-G(x, y) / k_\textrm{B} T} \right] , 
\end{equation}
where $G$ is the free energy function, $y$ is the variable integrated out of the profile, $x$ is the remaining variable, and $k_\textrm{B} T$ is the Boltzmann constant multiplied by the simulation temperature.

We calculate the free energy difference, $\Delta G$, by subtracting the free energy of the hybridized conformation from that of the solvent-facing conformation. To calculate the free energy of each state, we evaluate
\begin{equation} \label{eq:delG}
G_\textrm{state} = -k_\textrm{B} T \log \left[ \int_{x_\textrm{min}}^{x_\textrm{max}} \textrm{d}x \, e^{-G(x) / k_\textrm{B} T} \right] , 
\end{equation}
where $x_\textrm{min}$ and $x_\textrm{max}$ are the left and right boundaries of the state, defined by the flanking free energy maxima.


\section{Results and Discussion}

\subsection{Structural Order Parameters}

For each system, we calculate the probability distributions of the DNA order parameters
buckle, opening, propeller, rise, roll, shear, shift, slide, stagger, stretch, tilt, and twist
from unbiased molecular dynamics simulations.
These distributions (and the corresponding free energy profiles) give valuable information regarding the structures of the hybridized ground state and of the base pair of interest.
Imhof and Zahran ~\cite{imhof_effect_2013} calculated these distributions for the AT and GC matches as well as the GT mismatch.
For these base pairs, our results agree with theirs, despite our employing different DNA sequences.

\begin{figure}[h]
\centering
 \includegraphics[width=0.75\textwidth]{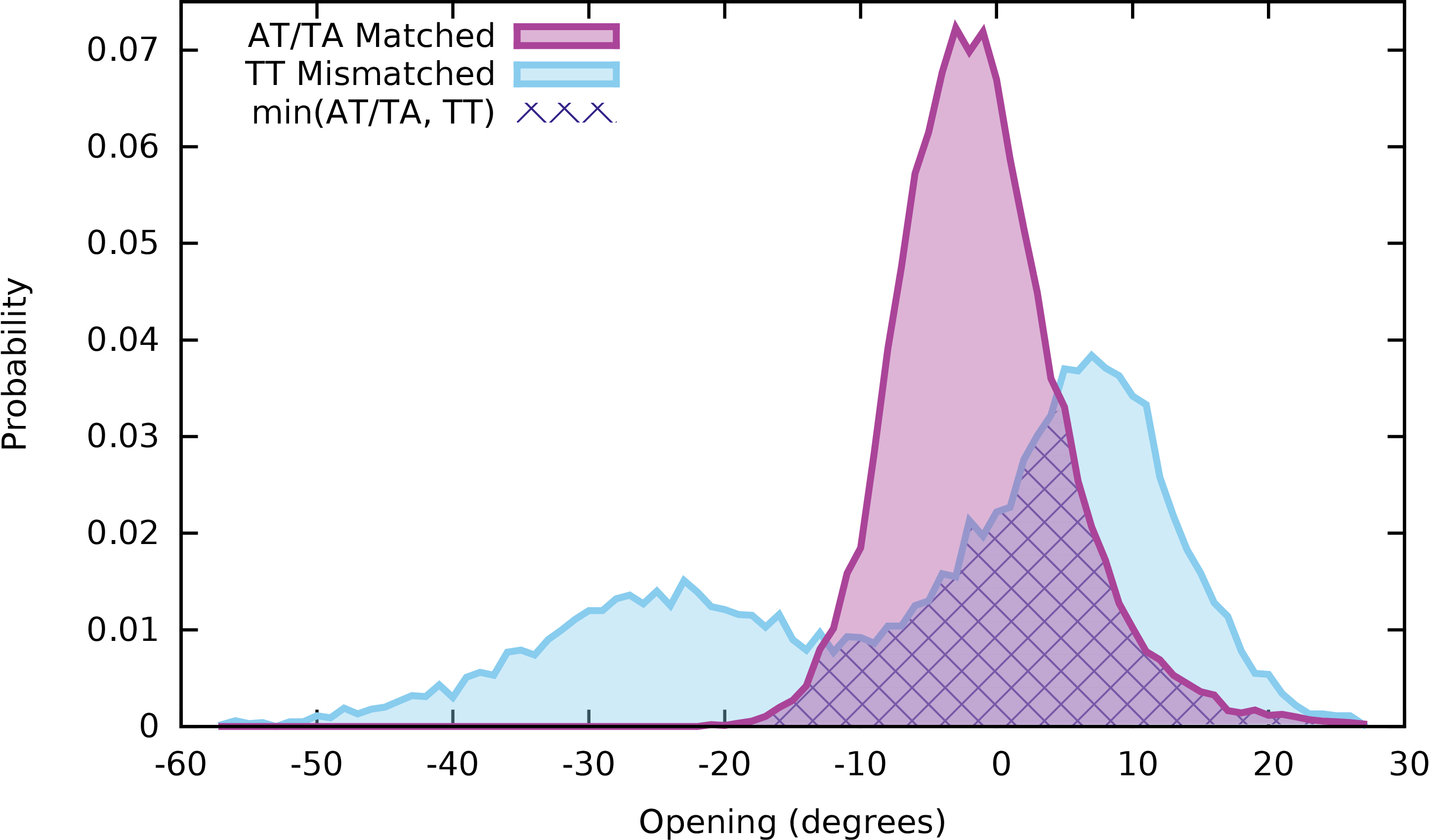}
 \caption{Probability overlap for opening.  The mean probability density of AT and TA matched base pairs (magenta) relative to the probability density of the TT mismatched system (blue).  The area in common (hatched) is the probability overlap; for this example it has a value of 0.47.}
 \label{fig:opening_min_area}
\end{figure}

For all studied base pairs, we measure the similarity in structure of matches and mismatches by calculating the probability overlap:
\begin{equation} \label{eq:minarea}
 \int_{-\infty}^{\infty} \mathrm{d}x \, \mathrm{min} \Bigl( P_{\mathrm{mis}}(x), P_{\mathrm{mat}}(x) \Bigr) ,
\end{equation}
which is the integrated area of the minimum of the mismatched probability density ($P_{\mathrm{mis}}$) and the mean of the corresponding matched probability densities ($P_{\mathrm{mat}}$).
This measure of overlap ranges from 0 to 1, where 0 indicates complete dissimilarity of structure and 1 indicates exact similarity.
We provide an illustrative example in Figure~\ref{fig:opening_min_area}.

\begin{table}[htbp]
 \begin{center}
 \newcommand{\mc}[3]{\multicolumn{#1}{#2}{#3}}
\begin{tabular}{l|c|c|c|c|c|c|c|c}\cline{2-9}
   & AA & CC & TT & AG & AC & CT & GT & \mc{1}{c|}{avg.}\\\hline
   \mc{1}{|l|}{stretch} & 0.00 & 0.03 & 0.01 & 0.00 & 0.18 & 0.05 & 0.09 & \mc{1}{c|}{0.05}\\\hline
   \mc{1}{|l|}{shear} & 0.00 & 0.02 & 0.00 & 0.76 & 0.00 & 0.64 & 0.00 & \mc{1}{c|}{0.20}\\\hline
   \mc{1}{|l|}{opening} & 0.35 & 0.38 & 0.47 & 0.21 & 0.53 & 0.41 & 0.72 & \mc{1}{c|}{0.44}\\\hline
   \mc{1}{|l|}{twist} & 0.29 & 0.34 & 0.23 & 0.91 & 0.46 & 0.96 & 0.44 & \mc{1}{c|}{0.52}\\\hline
   \mc{1}{|l|}{shift} & 0.54 & 0.45 & 0.86 & 0.59 & 0.64 & 0.67 & 0.77 & \mc{1}{c|}{0.65}\\\hline
   \mc{1}{|l|}{tilt} & 0.29 & 0.85 & 0.74 & 0.52 & 0.84 & 0.61 & 0.93 & \mc{1}{c|}{0.68}\\\hline
   \mc{1}{|l|}{slide} & 0.37 & 0.62 & 0.77 & 0.64 & 0.90 & 0.80 & 0.85 & \mc{1}{c|}{0.71}\\\hline
   \mc{1}{|l|}{stagger} & 0.50 & 0.78 & 0.79 & 0.56 & 0.80 & 0.71 & 0.80 & \mc{1}{c|}{0.71}\\\hline
   \mc{1}{|l|}{propeller} & 0.63 & 0.80 & 0.83 & 0.85 & 0.78 & 0.63 & 0.67 & \mc{1}{c|}{0.74}\\\hline
   \mc{1}{|l|}{buckle} & 0.73 & 0.76 & 0.93 & 0.88 & 0.86 & 0.78 & 0.94 & \mc{1}{c|}{0.84}\\\hline
   \mc{1}{|l|}{rise} & 0.85 & 0.70 & 0.77 & 0.91 & 0.93 & 0.92 & 0.95 & \mc{1}{c|}{0.86}\\\hline
   \mc{1}{|l|}{roll} & 0.86 & 0.87 & 0.95 & 0.84 & 0.84 & 0.89 & 0.93 & \mc{1}{c|}{0.88}\\\hline
   \mc{1}{|l|}{bend} & 0.93 & 0.92 & 0.95 & 0.77 & 0.88 & 0.92 & 0.95 & \mc{1}{c|}{0.90}\\\hline
   \mc{1}{|l|}{avg.} & 0.49 & 0.58 & 0.64 & 0.65 & 0.66 & 0.69 & 0.70 & \\\cline{1-8}
  \end{tabular}
 \end{center}
 \caption{Overlap between order parameter distributions of mismatched and corresponding matched base pairs, Eq. \ref{eq:minarea}, ordered both top-to-bottom and left-to-right by increasing overlap (as averaged over rows and columns, respectively). Values range from 0 to 1, where high numbers signify large similarity.}
 \label{tab:structural}
\end{table}

We calculate overlap values and present the results in Table~\ref{tab:structural}.
Our data demonstrate that, on average, GT is the most similar to its matched counterparts.
This is an unintuitive result, as GT is one of the easiest mismatches for MMR proteins to identify.\cite{kramer_different_1984, su_mispair_1988, thomas_heteroduplex_1991, kunkel_eukaryotic_2015}
In a similar vein, CC is one of the most difficult for MMR proteins to recognize,~\cite{kramer_different_1984, su_mispair_1988, thomas_heteroduplex_1991, kunkel_eukaryotic_2015} but it possesses one of the structures most dissimilar from matched configurations. This trend holds for bend, which has previously been determined to correlate with base flip,~\cite{sharma_differential_2014, sharma_dna_2013} and is thus a likely mechanism for MMR recognition.

Our data reveal that the distributions of stretch, shear, opening, twist, and shift vary the most when we compare mismatched with matched base pairs.
These order parameters are within the plane perpendicular to the DNA.
Order parameters presenting negligible differences include roll, rise, buckle, propeller, and stagger; all defined outside of the perpendicular plane.
These results indicate that replacing a matched base by a mismatched one distorts the DNA mostly within the plane of the base pair, not along the length of the molecule.
These distortions do not appear to correlate with MMR enzymes' ability to identify mismatches.~\cite{kramer_different_1984, su_mispair_1988, thomas_heteroduplex_1991, kunkel_eukaryotic_2015}

\subsection{Hydrogen Bonding}

We calculate the mean number of hydrogen bonds between the two bases of interest, X and Y.
We present our results in the second column of Table \ref{tab:isoHbond}.
As expected, the average number of hydrogen bonds for the matched pairs AT and TA is approximately two, and for pairs CG and GC it is approximately three. Mismatched base pairs contain a consistently lower number of hydrogen bonds than their matched counterparts. 
Because hydrogen bonding between opposing bases determines DNA stability,~\cite{	turner_free_1987,sponer_electronic_2001,kool_hydrogen_2001,yakovchuk_base-stacking_2006,khakshoor_measurement_2012,shankar_dna_2012} these results indicate decreased stability in the presence of mismatches; a result well-known experimentally.~\cite{peyret_nearest-neighbor_1999,santalucia_jr._thermodynamics_2004,harris_defects_2006,geyer_nucleobase_2003}

\begin{figure}
 \centering
 \includegraphics[height=0.4\textheight]{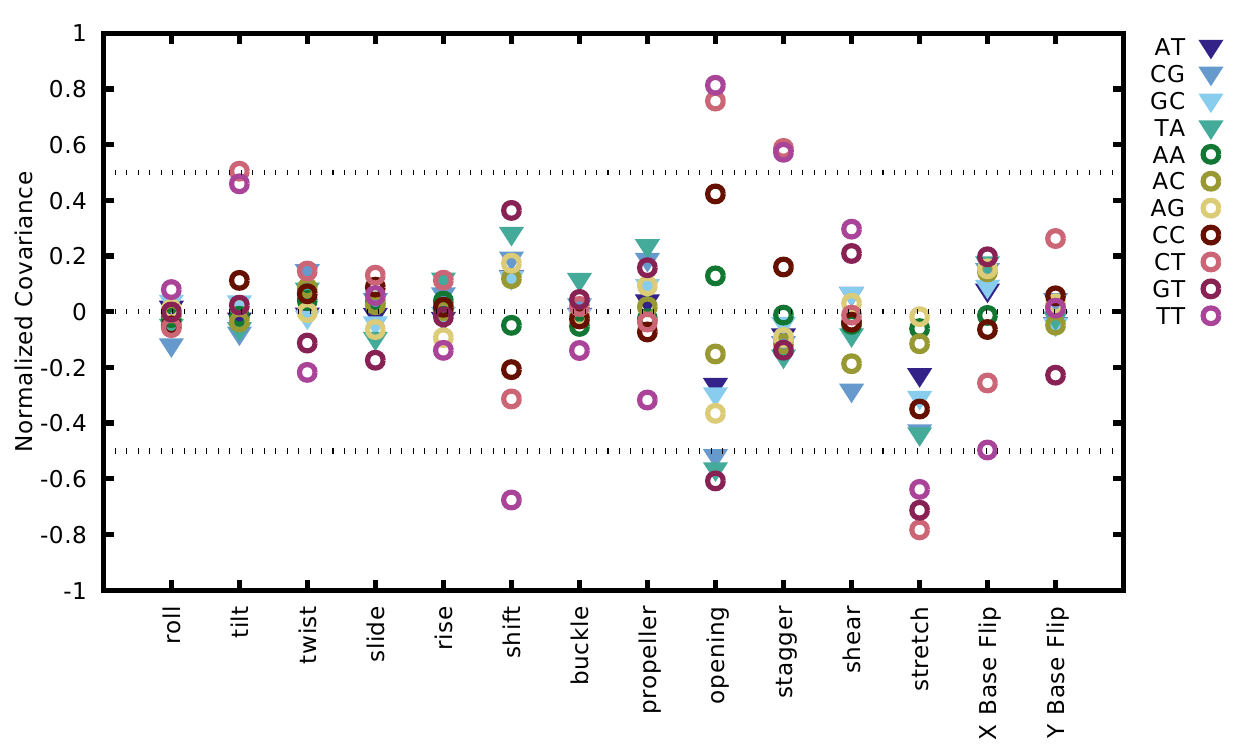}
 \caption{The normalized covariance between each structural parameter explored in this study and the number of hydrogen bonds.
	  Solid triangles correspond to matched base pairs; hollow circles correspond to mismatched pairs.
	  ``X Base Flip'' and ``Y Base Flip'' refer to the flip angles of bases X and Y, respectively.
	  }
 \label{fig:Hbondcorr}
\end{figure}

According to our results, hydrogen bonds do not systematically correlate with any of the analyzed structural parameters (Fig.~\ref{fig:Hbondcorr}), though weak trends are discernible. The correlation with stretch, if present, is always negative. This is an expected result; as bases drift from one another, their ability to hydrogen bond decreases.
Interestingly, the structures of mismatched pairs CT, GT, and TT correlate more with hydrogen bonding than any other pair.

\begin{table}[htbp]
{
\newcommand{\mc}[3]{\multicolumn{#1}{#2}{#3}}
\begin{center}
\begin{tabular}{cc|cc|ccc|}\cline{3-7}
 &  & \mc{2}{c|}{Unbiased MD} & \mc{3}{c|}{Metadynamics} \\
 &  & Time & Mean & Time & $\Delta G_X$ & $\Delta G_Y$ \\
 &  & (ns) & \# Hbonds & (ns) & (kJ/mol) & (kJ/mol) \\\hline
\mc{1}{|l|}{\parbox[t]{2mm}{\multirow{4}{*}{\rotatebox[origin=c]{90}{Matched}}}} & AT & 100 & 1.99 & 700 & 29 & 26\\
\mc{1}{|l|}{} & TA & 100 & 1.95 & 700 & 28 & 71\\
\mc{1}{|l|}{} & CG & 100 & 2.97 & 700 & 22 & 68\\
\mc{1}{|l|}{} & GC & 100 & 2.98 & 700 & 11 & 7\\\hline
\mc{1}{|l|}{\parbox[t]{2mm}{\multirow{7}{*}{\rotatebox[origin=c]{90}{Mismatched}}}} & AA & 100 & 1.01 & 1000 & 40 & 60\\
\mc{1}{|l|}{} & AC & 100 & 1.96 & 700 & -1 & -1\\
\mc{1}{|l|}{} & AG & 100 & 1.08 & 700 & 19 & 56\\
\mc{1}{|l|}{} & CC & 100 & 1.08 & 700 & 8 & 8\\
\mc{1}{|l|}{} & CT & 100 & 2.41 & 700 & 23 & 16\\
\mc{1}{|l|}{} & GT & 100 & 1.97 & 700 & 8 & 21\\
\mc{1}{|l|}{} & TT & 100 & 1.62 & 700 & 10 & 3\\\hline
\end{tabular}
\end{center}
}
  \caption{First column: the time (in nanoseconds) of unbiased simulation.
	   Second column: The mean number of hydrogen bonds between the bases of interest during unbiased simulation.
	   Third column: the time (in nanoseconds) of biased metadynamics simulation.
	   Fourth and fifth columns: The free energy of base flip for the first and second (X and Y of XY, respectively) base of each pair.
	   }
 \label{tab:isoHbond}
\end{table}

\subsection{Base Flip}

Unbiased simulations are insufficient to determine the free energy of base flip.
The free energy barrier to flip out of the DNA double helix and into solution is too immense to overcome on the simulation time scale.
To ameliorate this complication, we applied the well-tempered metadynamics algorithm~\cite{laio_metadynamics:_2008,barducci_well-tempered_2008} which computes the free energy surfaces (FES) of base flip over the entire domain.
This technique continually adds Gaussian-shaped biasing potentials to each reaction coordinate, driving the system toward previously unexplored configurations.

\begin{figure}[htbp]
\centering
 \includegraphics[width=\textwidth]{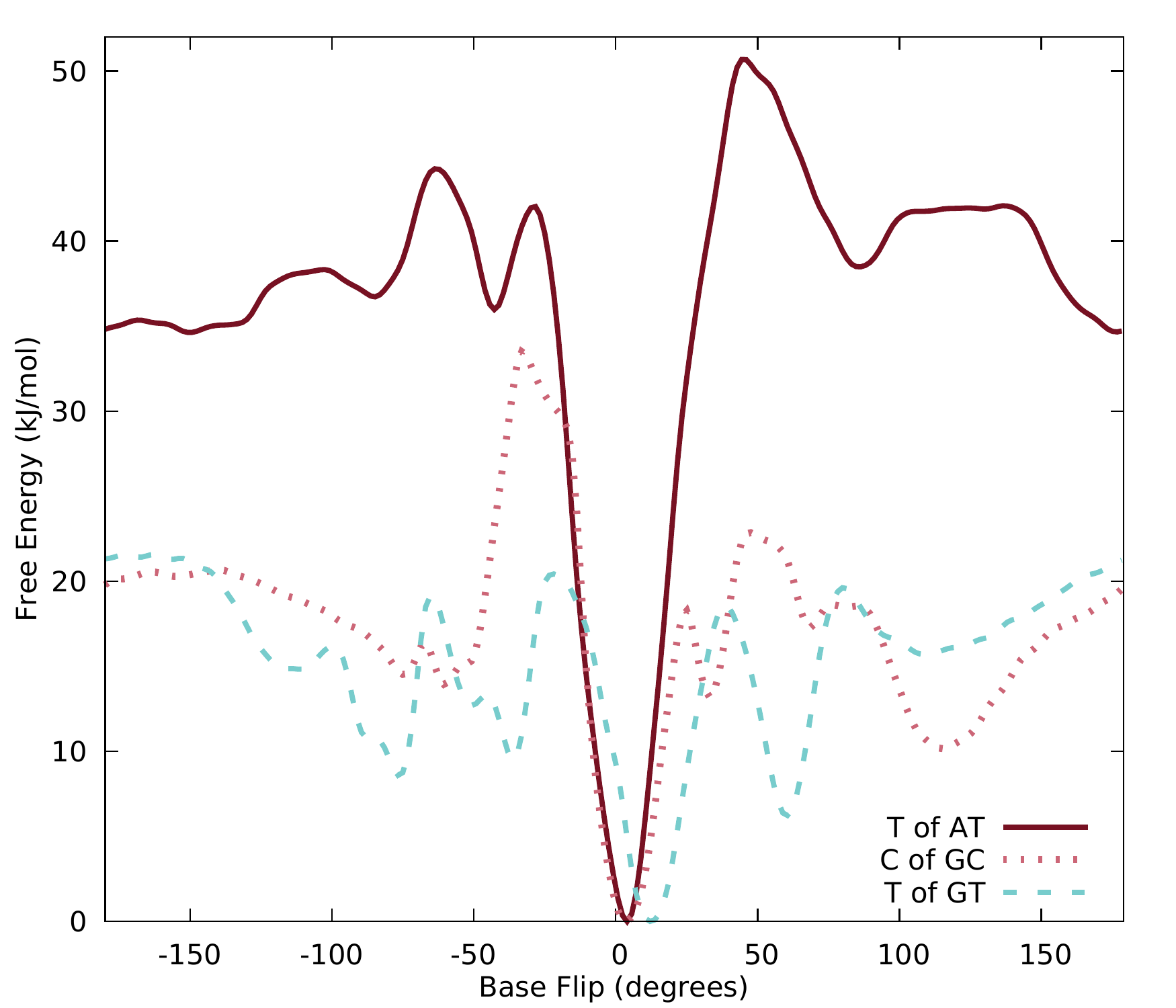}
 \caption{
The free energy of flipping the T base out of a mismatched GT pair (dashed blue), compared to the matched counterparts AT (solid red) and GC (dashed red). The mismatched GT pair shows a slight shift in the global minimum.
 }
 \label{fig:baseflip1D}
\end{figure}

We compute an FES of base flip for each matched and mismatched pair. 
We illustrate a representative result in Figure~\ref{fig:baseflip1D}, which compares the free energies of flipping a T base in a GT mismatch with those of the corresponding T base in an AT matched pair and the C base of a matched GC pair.
We calculate these one-dimensional profiles from 2-CV metadynamics simulations by Eq.~\ref{eq:2Dto1D}.
Each of these bases possess the same local environment as the others.
For each system, we identify a global minimum in the FES corresponding to the hybridized ground state. 
Extrahelical conformations form local minima separated from the principal minimum by activation barriers.
Similar results have been reported in the literature for a subset of the base pairs considered in this study~\cite{banavali_free_2002, imhof_effect_2013, yin_dynamics_2014}.

\begin{figure}[h]
 \centering
 \includegraphics[height=0.35\textheight]{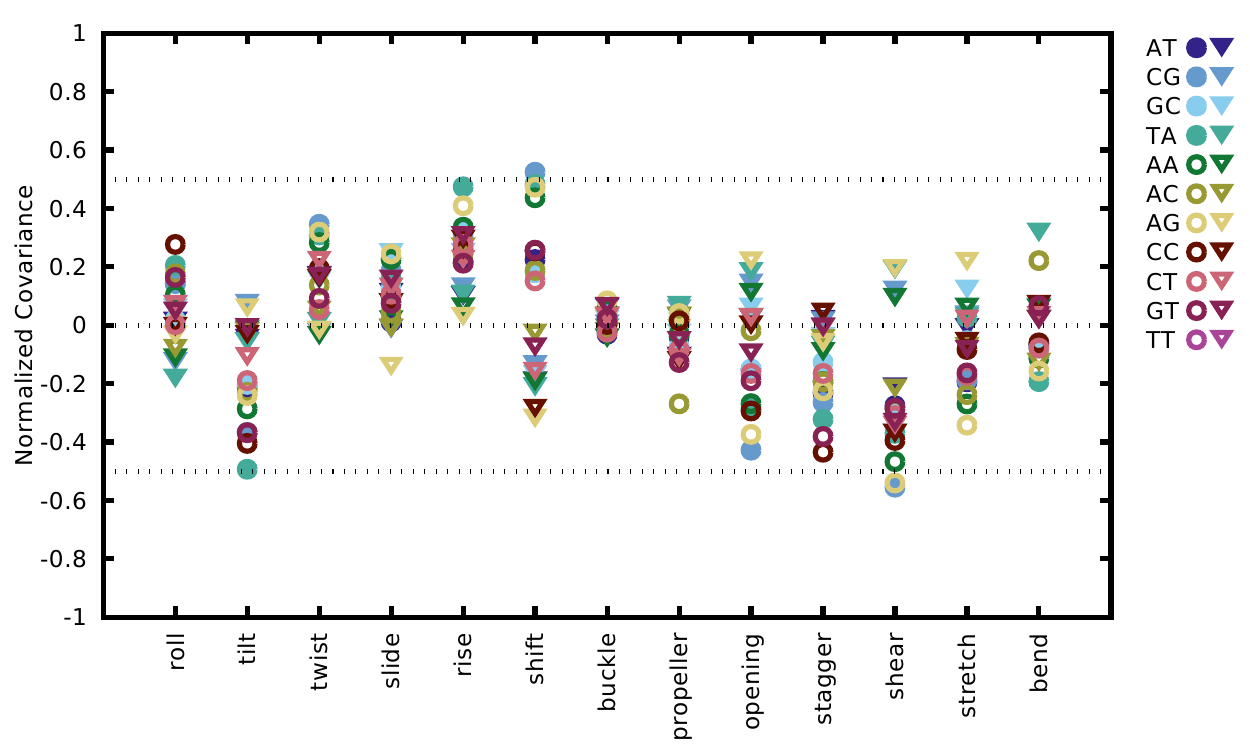}
 \caption{The normalized covariance between base flip and each structural order parameter explored in this study,
	  over the entire base flip reaction coordinate.
	  Solid shapes represent matched pairs; hollow shapes represent mismatched pairs. Circles represent the first base of a pair (e.g. ``A'' of ``AT''); triangles are the second base.
	  }
 \label{fig:bfcov}
\end{figure}

Base flip may resemble other order parameters when a base rotates into the DNA polymer;
indeed, it correlates weakly with twist and shear in unbiased simulation.
Once the base rotates out of the DNA strand it does not correlate with any of the other order parameters. (Fig. \ref{fig:bfcov})
We therefore conclude that base flip is a useful descriptor of base pair structure, building on the thirteen order parameters listed in Table~\ref{tab:structural}.
Despite no significant correlations, slight trends appear. 
For most structural parameters, the first and second bases of a pair have opposing correlations. For example, the parameter ``shift'' correlates positively with the base flip of the first base and negatively with the second. This is consistent with the opposing symmetries of hybridized DNA strands.

Calculations of base flip free energy profiles are difficult to converge due to long relaxation timescales;
Imhof et. al.~\cite{imhof_effect_2013} observed a notable ($\sim$10$^\circ$) shift of the minimum from unbiased to biased simulation of the TG mismatched system.
They hypothesize that the long timescale of flipping and re-stacking causes the free energy to converge slowly.
We notice a similar shift when testing their base flip reaction coordinate, first defined by MacKerell et. al.,~\cite{banavali_free_2002} with 50~ns metadynamics simulations and biasing only one base, but no such shift when adopting the updated base flip coordinate~\cite{song_improved_2009} with a bias potential applied to two bases.
To differentiate if this results from the definition of base flip or from the number of biased coordinates, we performed a simulation biasing only one base with the updated base flip coordinate and found still no shift of the minimum.
This would indicate that the choice of base flip coordinate is crucial not only for accuracy, as discussed in ref.~\cite{song_improved_2009}, but also for convergence.

\begin{figure}[htbp]
 \centering
 \begin{subfigure}[t]{0.45\textwidth}
  \includegraphics[width=\textwidth]{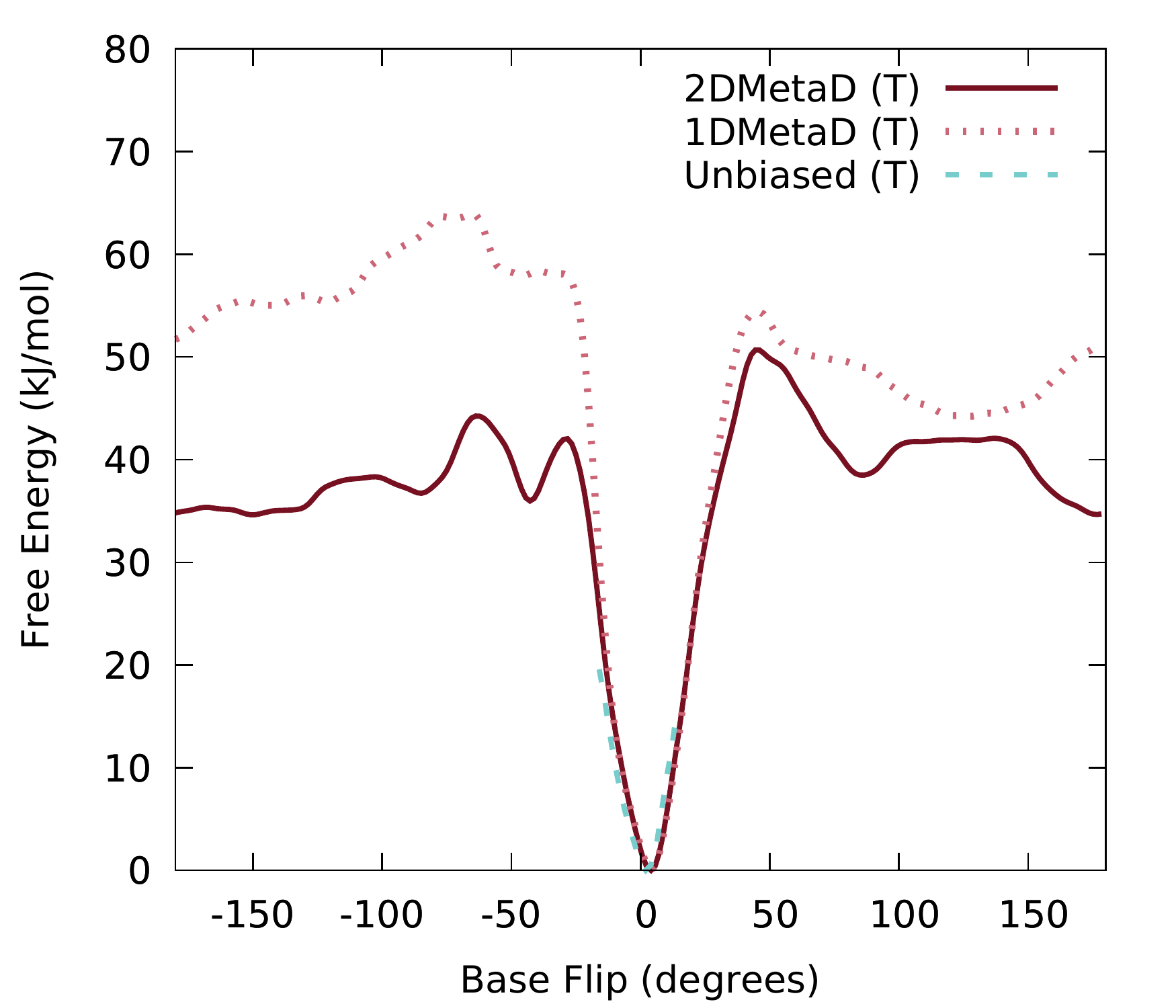}
  \caption{AT}
 \end{subfigure} \quad
 \begin{subfigure}[t]{0.45\textwidth}
  \includegraphics[width=\textwidth]{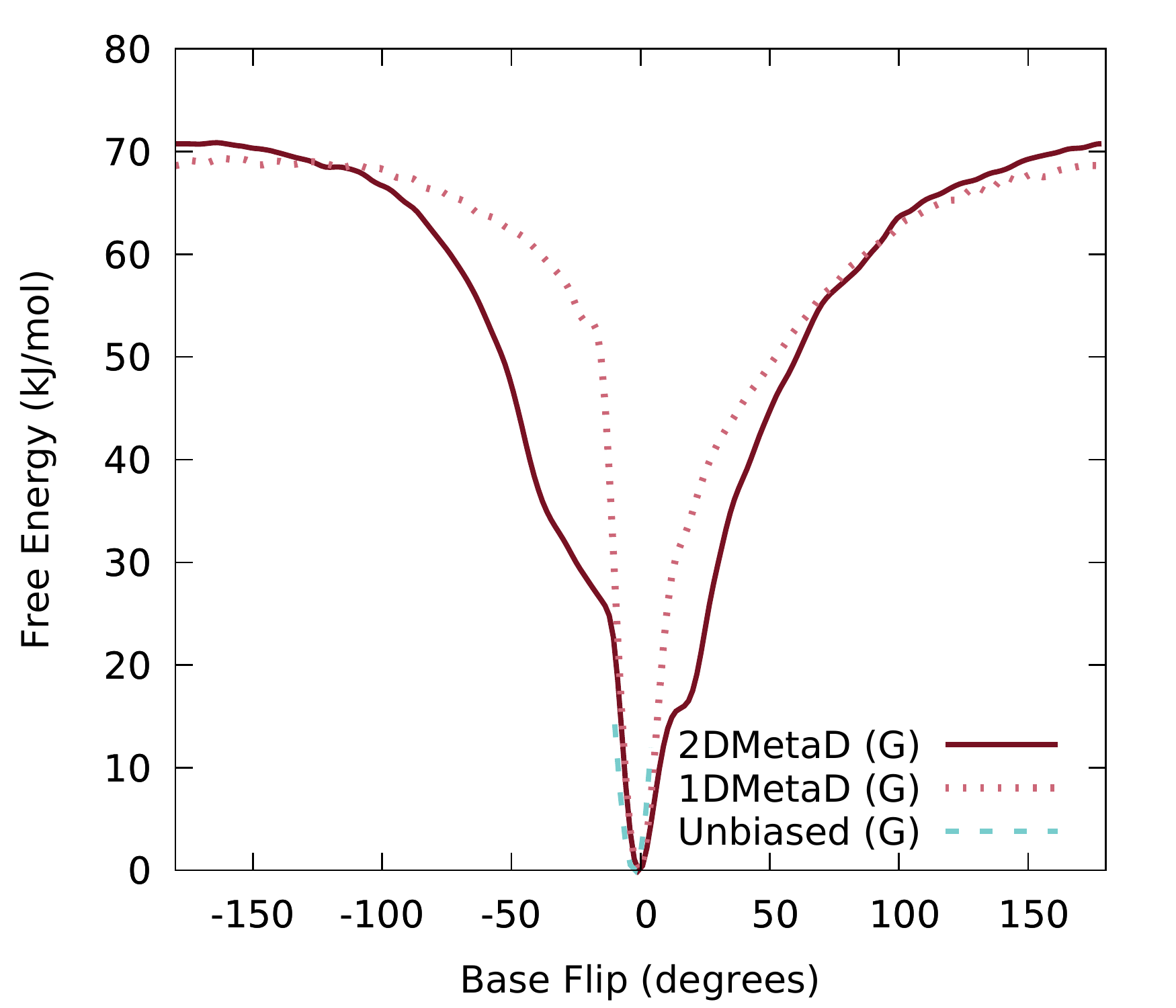}
  \caption{CG}
 \end{subfigure}

 \begin{subfigure}[t]{0.45\textwidth}
  \includegraphics[width=\textwidth]{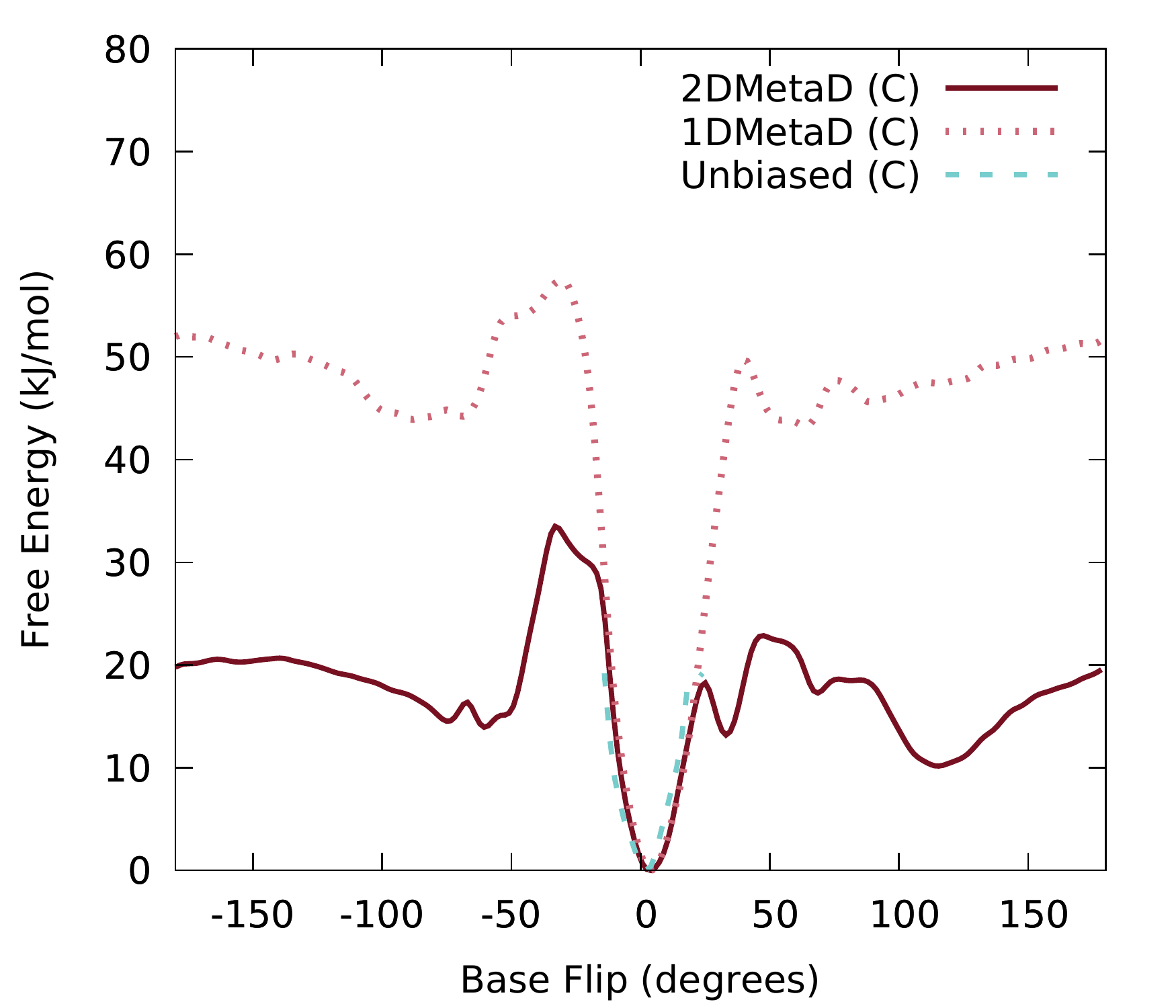}
  \caption{GC}
 \end{subfigure} \quad
  \begin{subfigure}[t]{0.45\textwidth}
   \includegraphics[width=\textwidth]{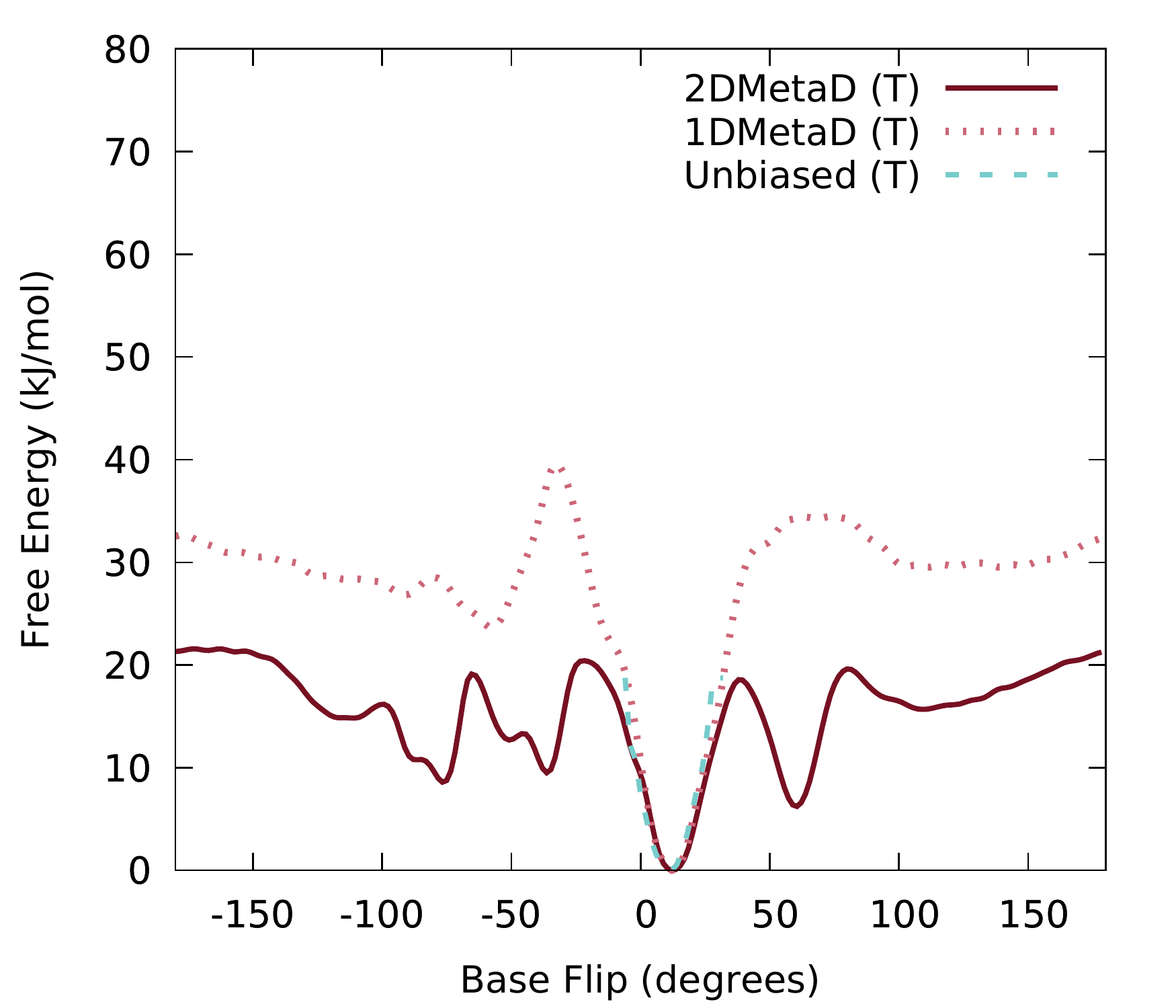}
   \caption{GT}
  \end{subfigure}
 \caption{Comparison of base flip free energy profiles for four systems: (a) T of an AT pair, (b) G of a CG pair, (c) C of a GC pair, and (d) T of a GT pair.
 	Shown are results from unbiased molecular dynamics simulations (dashed blue), metadynamics simulations in which only the flip angle of the T base was biased (dashed red), and metadynamics simulations in which the angles of both bases were biased (solid red).
 	The 1D free energy profiles are consistently greater than or equal to their 2D counterparts, indicating the 2D biasing potentials are more accurate.
	  }
 \label{fig:1CVbad}
\end{figure}

Simulations which bias two bases exhibit lower free energies than those which bias only one base. (Fig.~\ref{fig:1CVbad})
Because a lower free energy is favorable, this indicates the 2D biasing potentials possess greater accuracy than their 1D counterparts.
The 2D potentials allow the system to explore other favorable states, and better represent the probability landscape of the force field.
With our 2D potential we identify configurations we would otherwise not observe biasing only one base;
this explains why our results disagree with previous studies: they each bias one base where we bias two.~\cite{banavali_free_2002, onei_structures_2008, song_improved_2009, imhof_effect_2013, law_base-flipping_2011}

\begin{figure}[htbp]
 \centering
 \begin{subfigure}[t]{0.45\textwidth}
  \includegraphics[width=\textwidth]{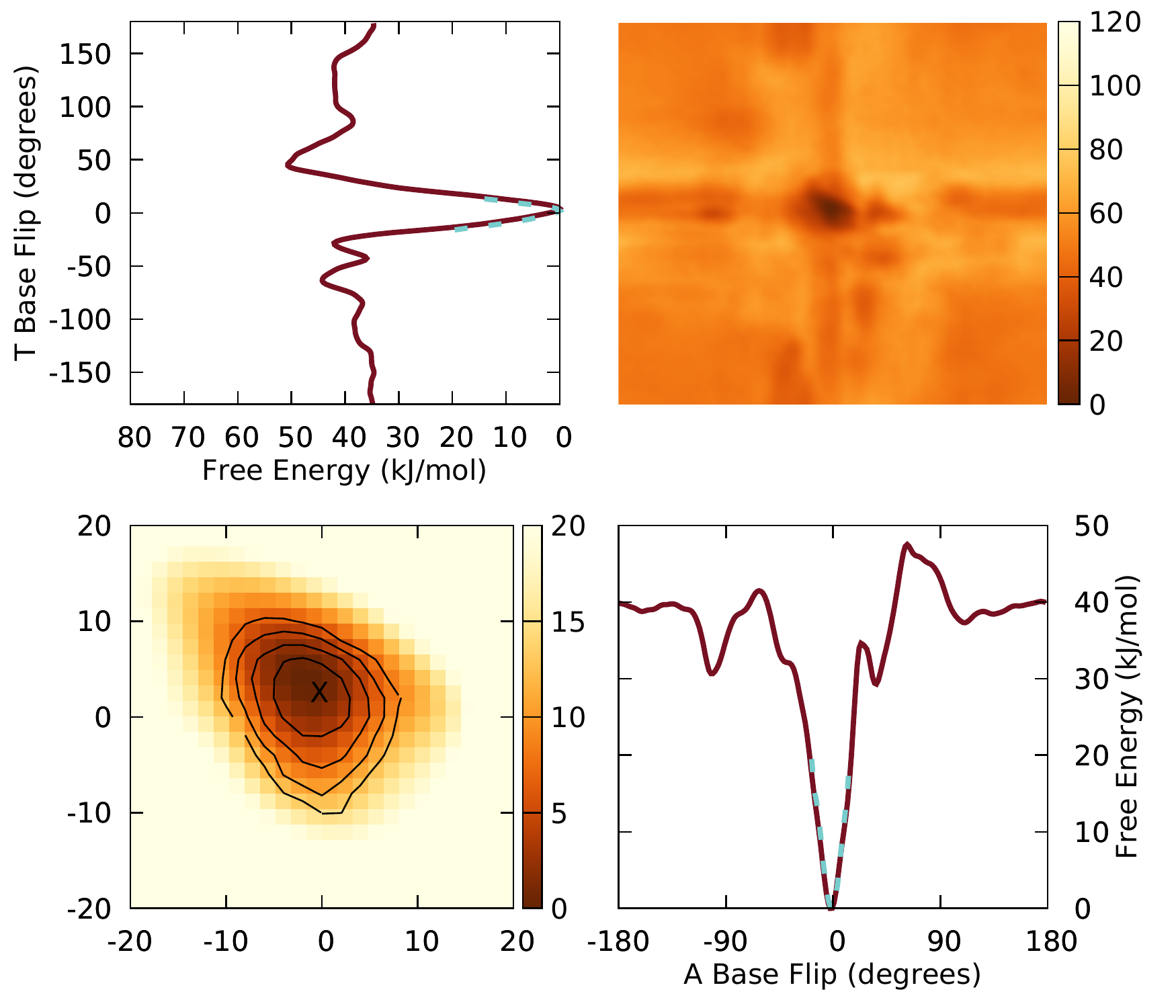}
  \caption{AT}
 \end{subfigure} \quad
 \begin{subfigure}[t]{0.45\textwidth}
  \includegraphics[width=\textwidth]{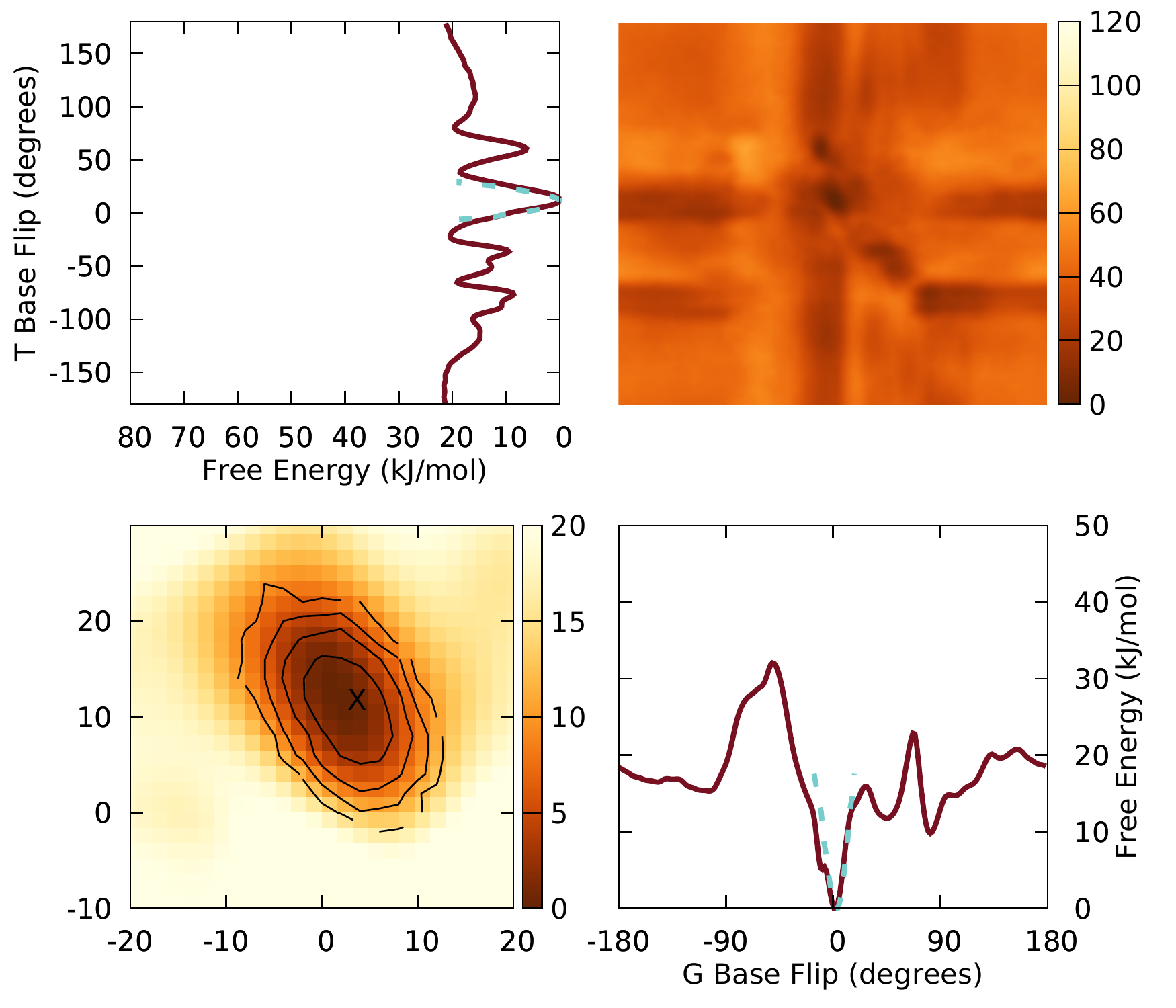}
  \caption{GT}
 \end{subfigure}

 \begin{subfigure}[t]{0.45\textwidth}
  \includegraphics[width=\textwidth]{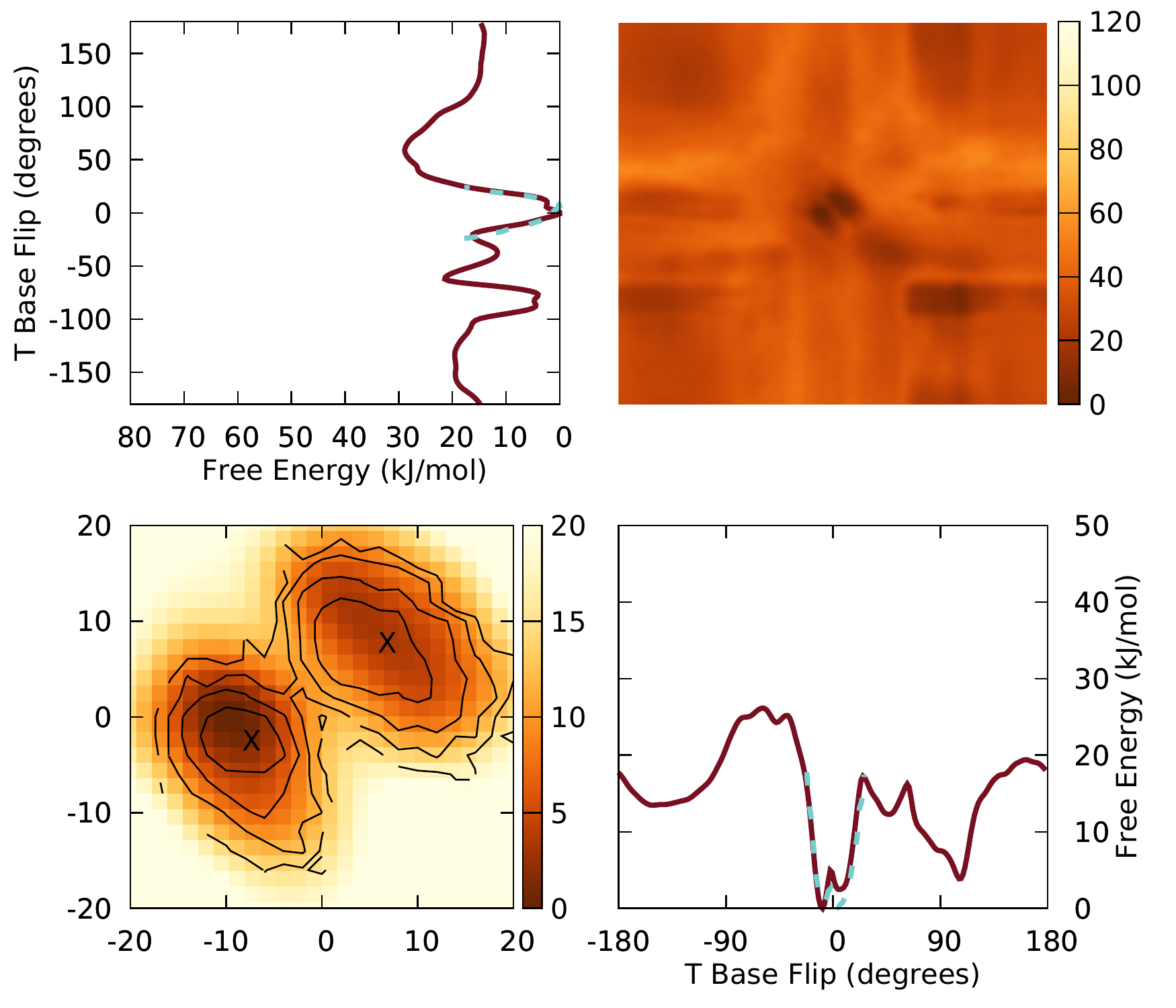}
  \caption{TT}
 \end{subfigure} \quad
  \begin{subfigure}[t]{0.45\textwidth}
   \includegraphics[width=\textwidth]{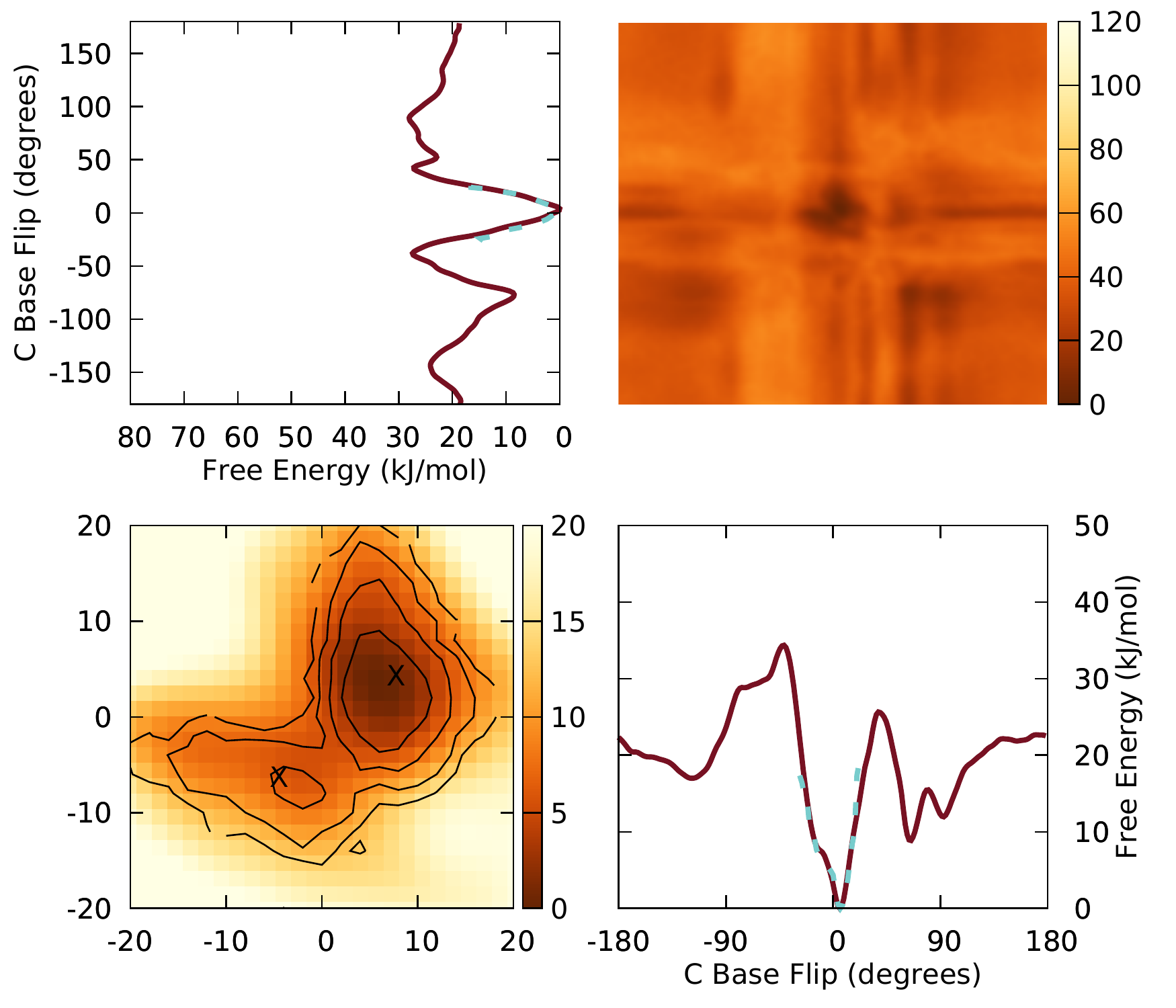}
   \caption{CC}
  \end{subfigure}
 \caption{Free energy of base flip for the matched pair (a) AT and (b-d) the mismatched pairs GT, TT, and CC.
	Shown in the upper right panel of each sub-figure is the two-dimensional free energy surface as a function of the base flip angles of both bases, as computed from metadynamics simulations. The top left and bottom right panels depict one-dimensional projections of this surface onto the base flip angle of the first (top left) and second (bottom right) base, together with results from unbiased simulations (dashed line). The bottom left panel grants a magnified view of the primary minimum in the two-dimensional free energy surface to 20 kJ/mol (heat map), together with results from unbiased simulations (contour lines). Flip angles demarcated by crosses correspond to local minimum configurations, as presented in Fig.~\ref{fig:mulmin}. Notably, a secondary minimum appears in multiple simulations at approximately (90$^\circ$, -100$^\circ$). This corresponds to the configuration presented by Fig.~\ref{fig:secondary_minimum}.
	}
 \label{fig:baseflip2D}
\end{figure}

\begin{figure}[htbp]
 \centering
 \begin{subfigure}[t]{0.45\textwidth}
  \includegraphics[width=\textwidth]{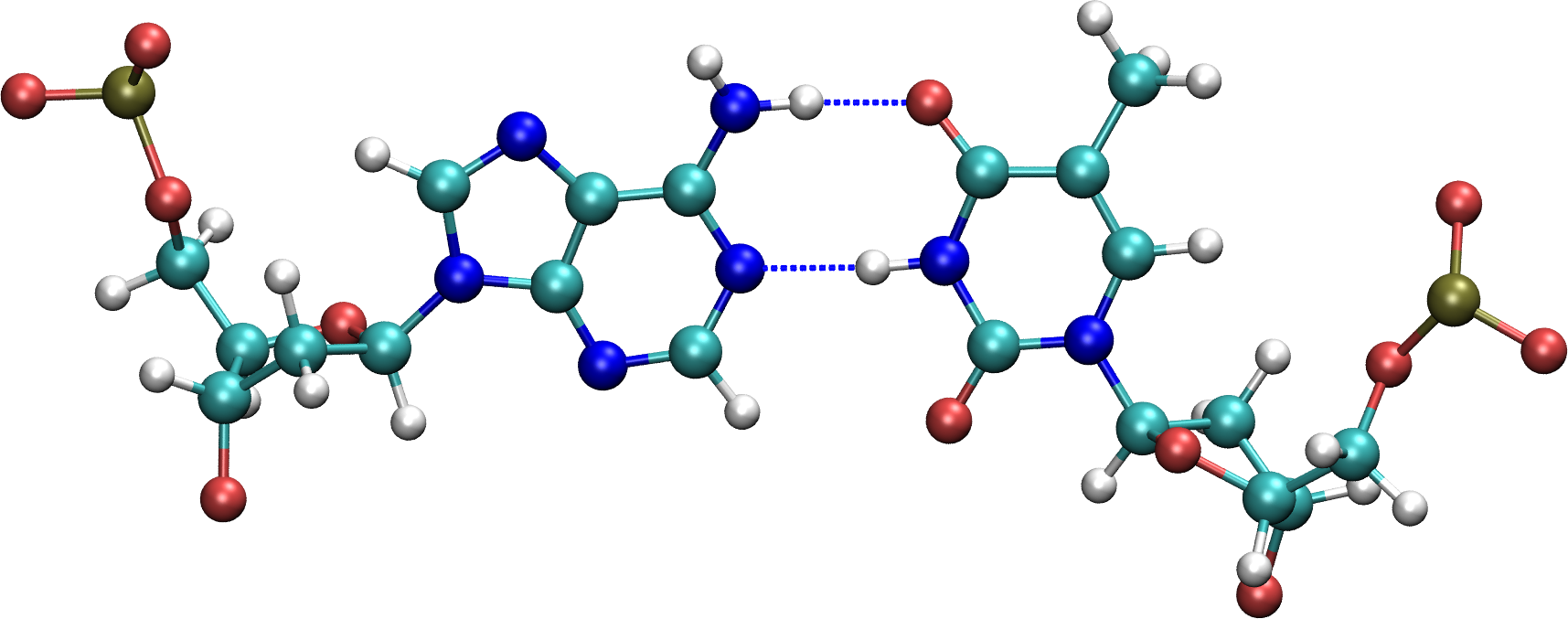}
  \caption{AT minimum}
 \end{subfigure}
 \quad
 \begin{subfigure}[t]{0.45\textwidth}
  \includegraphics[width=\textwidth]{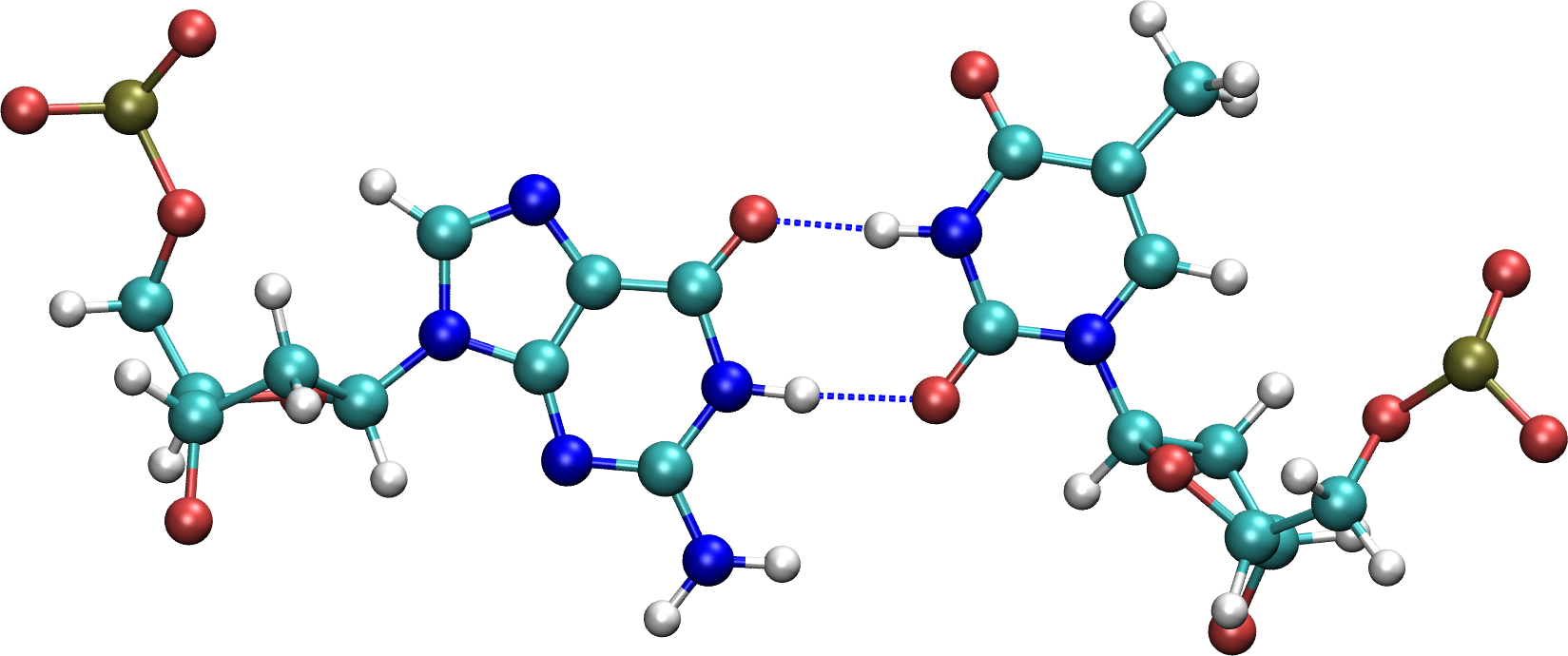}
  \caption{GT minimum}
 \end{subfigure}
 
 \begin{subfigure}[t]{0.45\textwidth}
  \includegraphics[width=\textwidth]{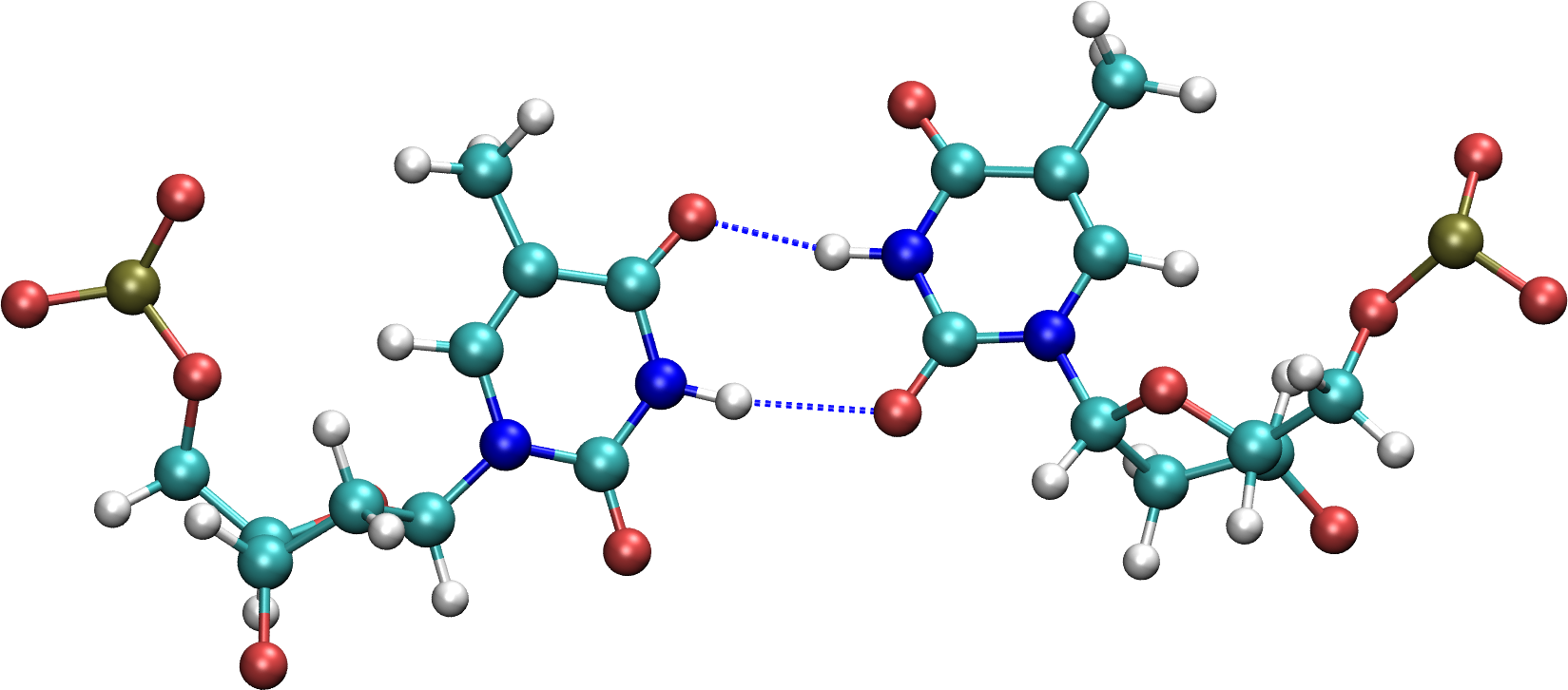}
  \caption{TT primary minimum ($\sim$52\%)}
 \end{subfigure}
 \quad
 \begin{subfigure}[t]{0.45\textwidth}
  \includegraphics[width=\textwidth]{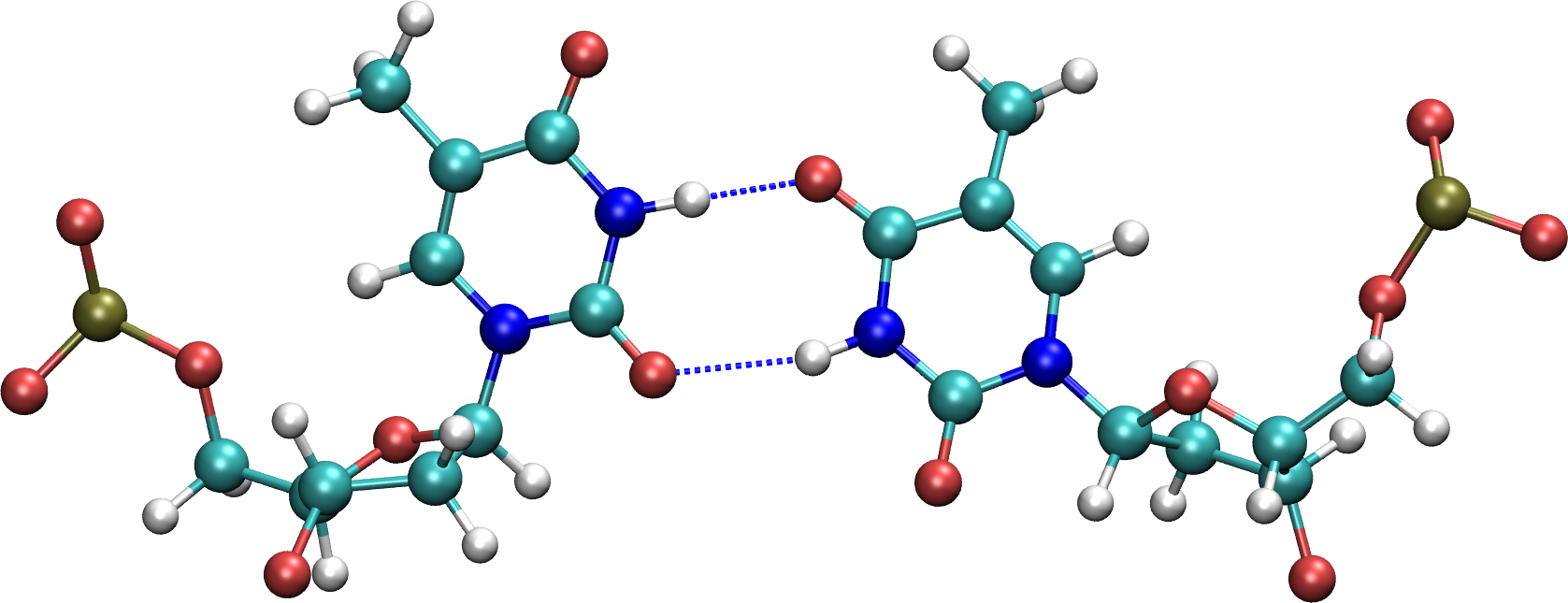}
  \caption{TT secondary minimum ($\sim$48\%)}
 \end{subfigure}
 
 \begin{subfigure}[t]{0.45\textwidth}
  \includegraphics[width=\textwidth]{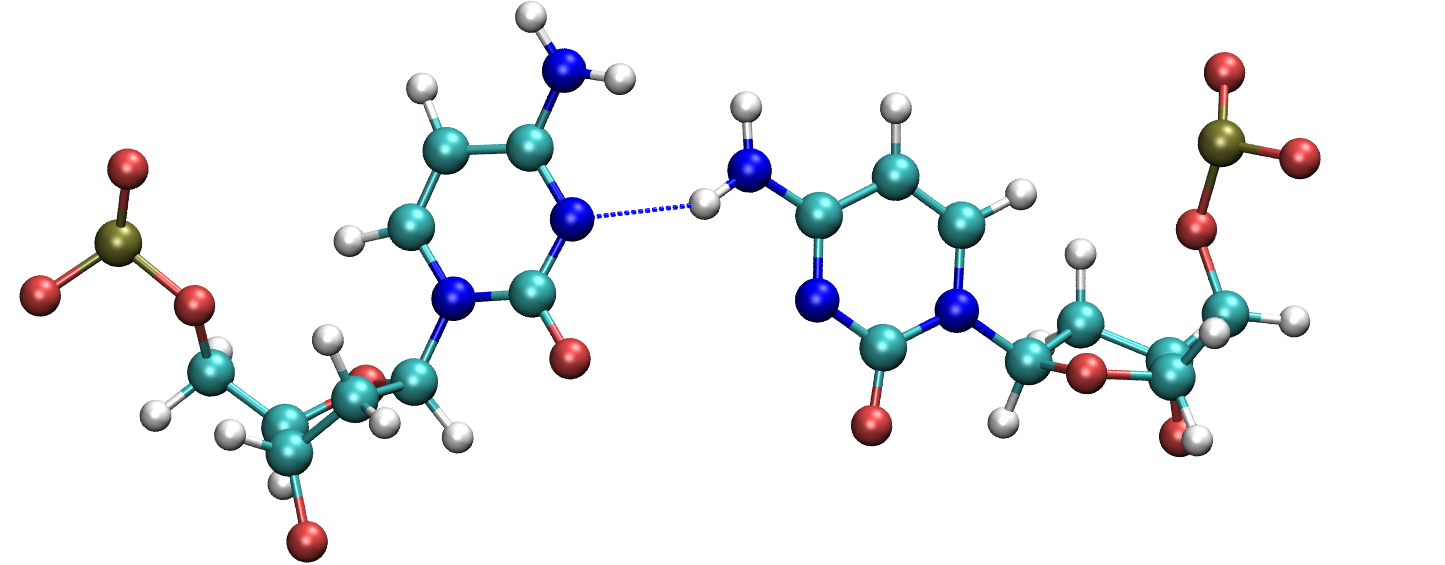}
  \caption{CC primary minimum ($\sim$51\%)}
 \end{subfigure}
 \quad
 \begin{subfigure}[t]{0.45\textwidth}
  \includegraphics[width=\textwidth]{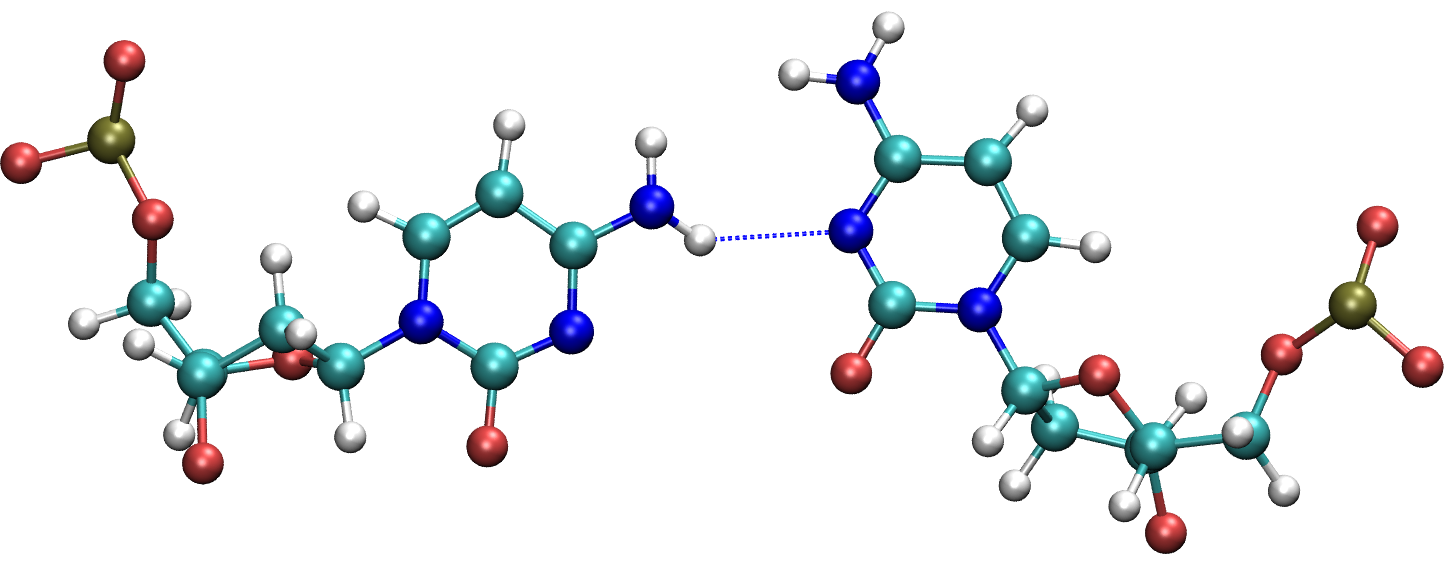}
  \caption{CC secondary minimum ($\sim$49\%)}
 \end{subfigure}
 \caption{Typical wobble structures for AT, GT, TT, and CC pairs as determined by clustering analysis.
          Hydrogen bonds as identified by the 3DNA software~\cite{lu_3dna:_2008} are expressed as dashed lines.
          The base flip angles of these configurations are presented in Figure~\ref{fig:baseflip2D}.
          }
 \label{fig:mulmin}
\end{figure}

\begin{figure}[htbp]
\centering
 \begin{center}
  \includegraphics[height=0.6\textheight]{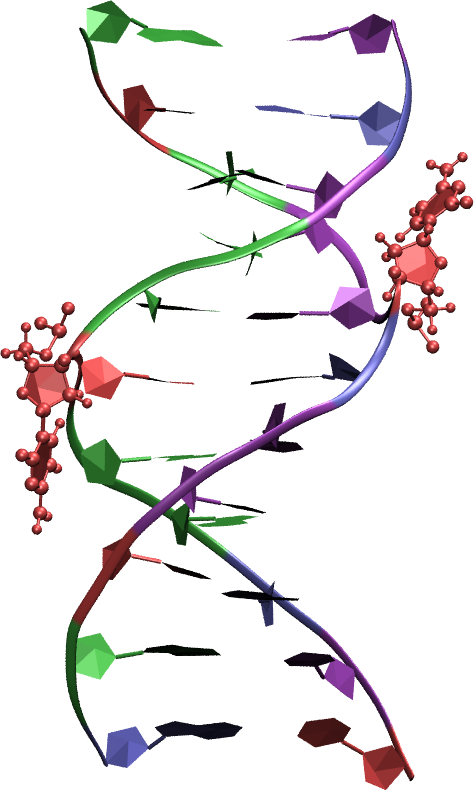}
 \end{center}
 \caption{
      Typical configuration of the secondary minimum (e-motif) for the CC pair as demonstrated in Figure~\ref{fig:baseflip2D}. Both bases of interest are no longer stacked within the DNA polymer, but are flipped to be solvent-facing. The bases lie alongside the DNA, within the minor groove, and pointing toward the 5' termini.
      }
 \label{fig:secondary_minimum}
\end{figure}

\begin{figure}[h]
 \centering
 \begin{subfigure}[t]{0.45\textwidth}
  \includegraphics[width=\textwidth]{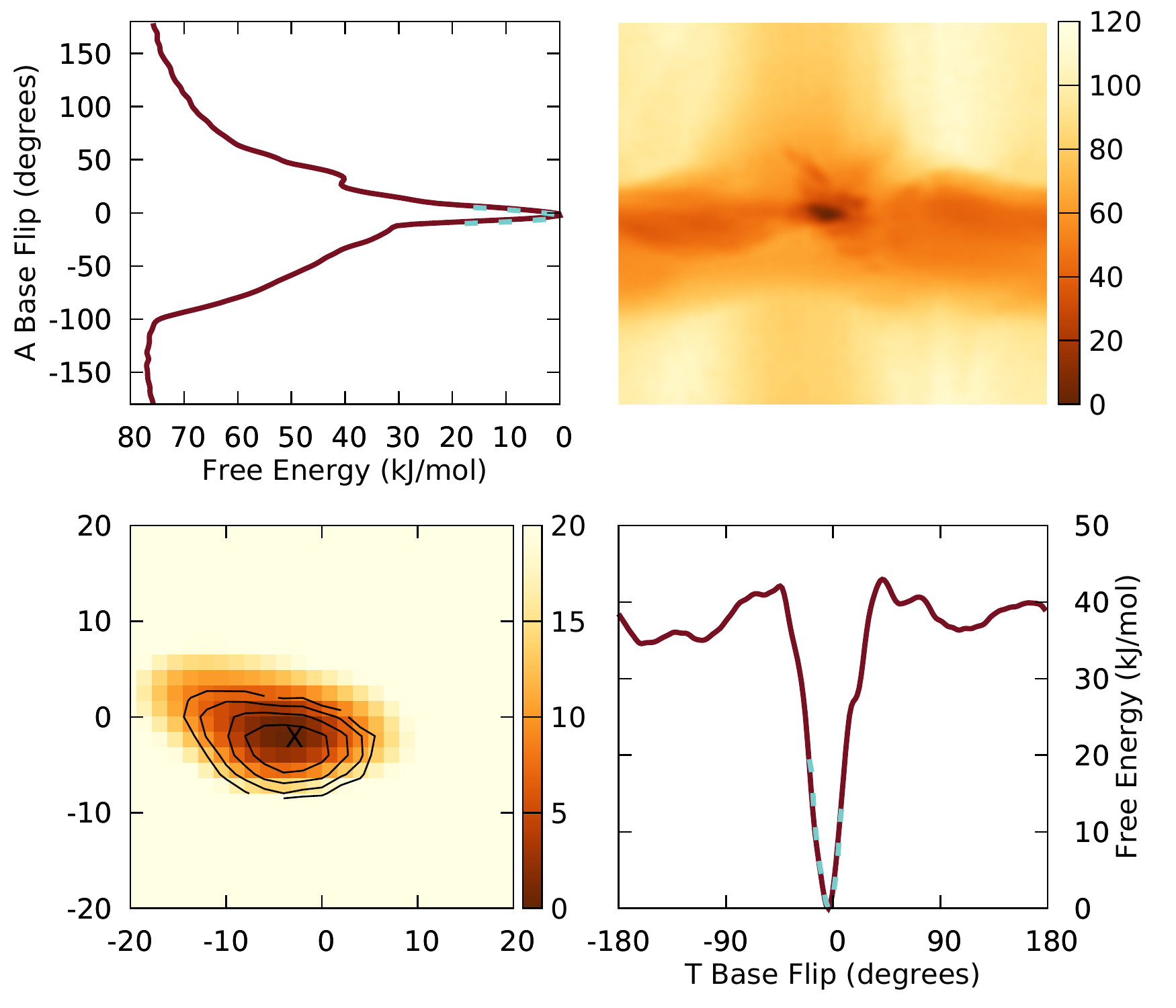}
  \caption{TA}
 \end{subfigure} \quad
 \begin{subfigure}[t]{0.45\textwidth}
  \includegraphics[width=\textwidth]{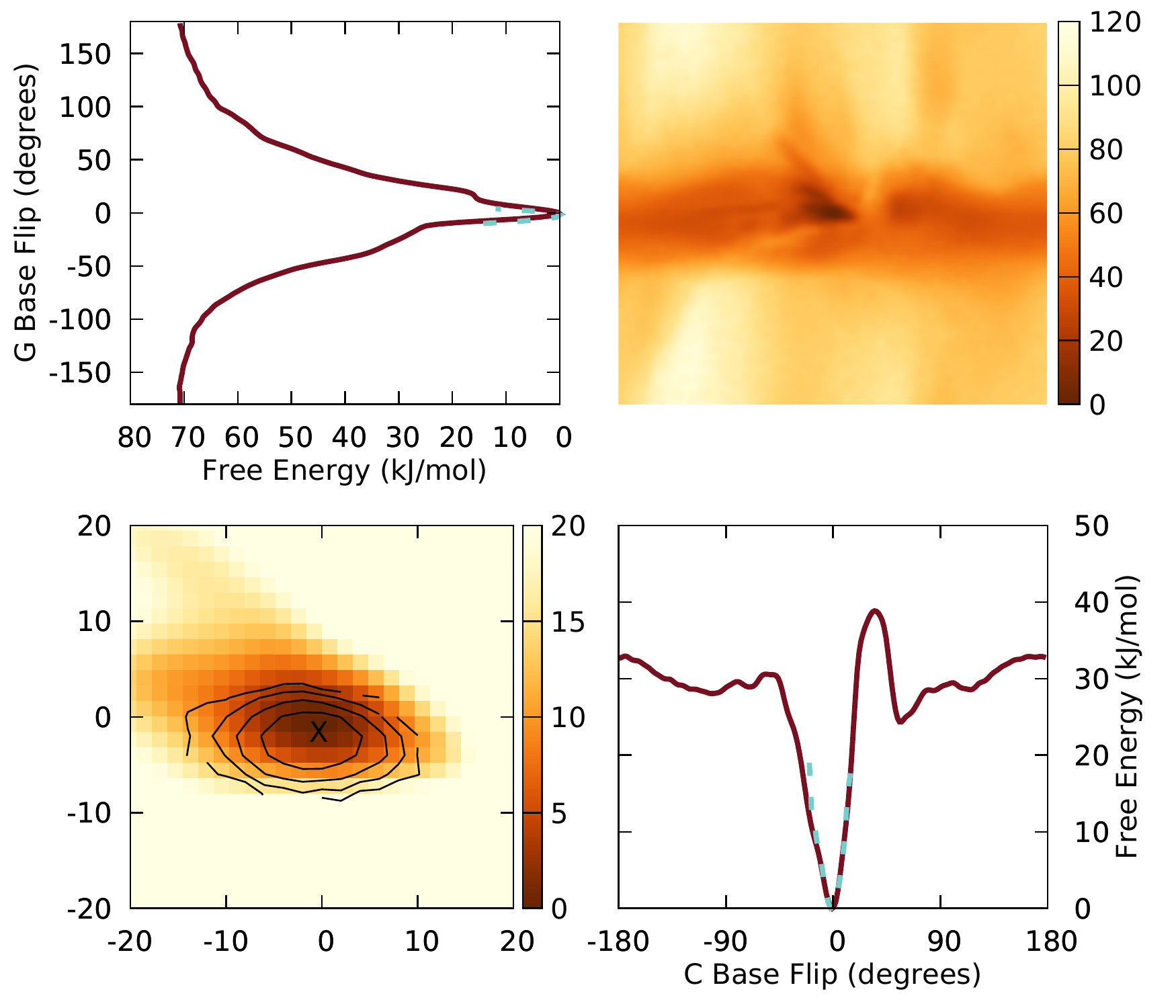}
  \caption{CG}
 \end{subfigure}

 \begin{subfigure}[t]{0.45\textwidth}
  \includegraphics[width=\textwidth]{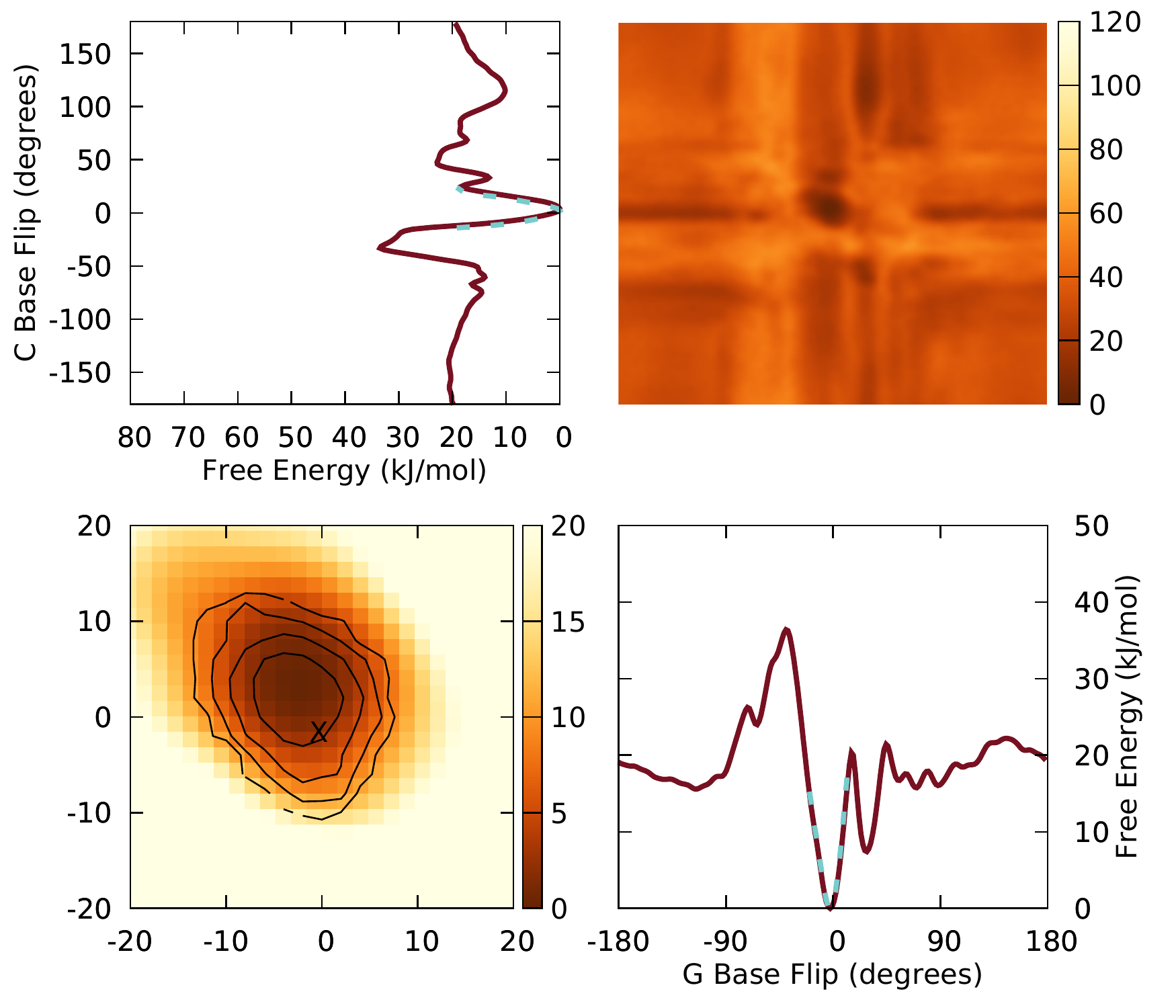}
  \caption{GC}
 \end{subfigure} \quad
 \caption{\label{fig:baseflipTACGGC}
Free energy of base flip for the matched pairs (a) TA, (b) CG, and (c) GC. Figure~\ref{fig:baseflip2D} describes each sub-figure.}
\end{figure}

\begin{figure}[h]
 \centering
 \begin{subfigure}[t]{0.45\textwidth}
  \includegraphics[width=\textwidth]{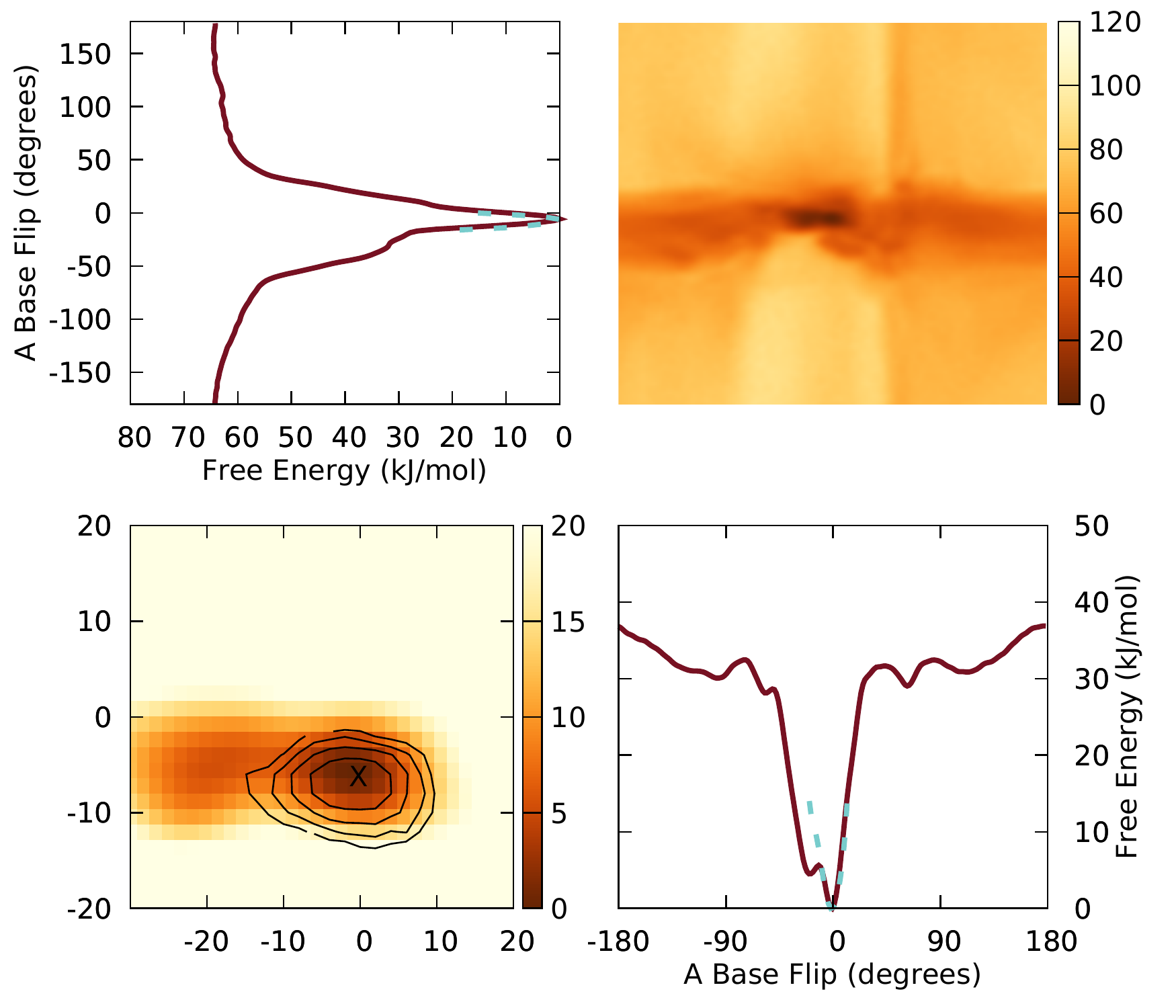}
  \caption{AA}
 \end{subfigure} \quad
 \begin{subfigure}[t]{0.45\textwidth}
  \includegraphics[width=\textwidth]{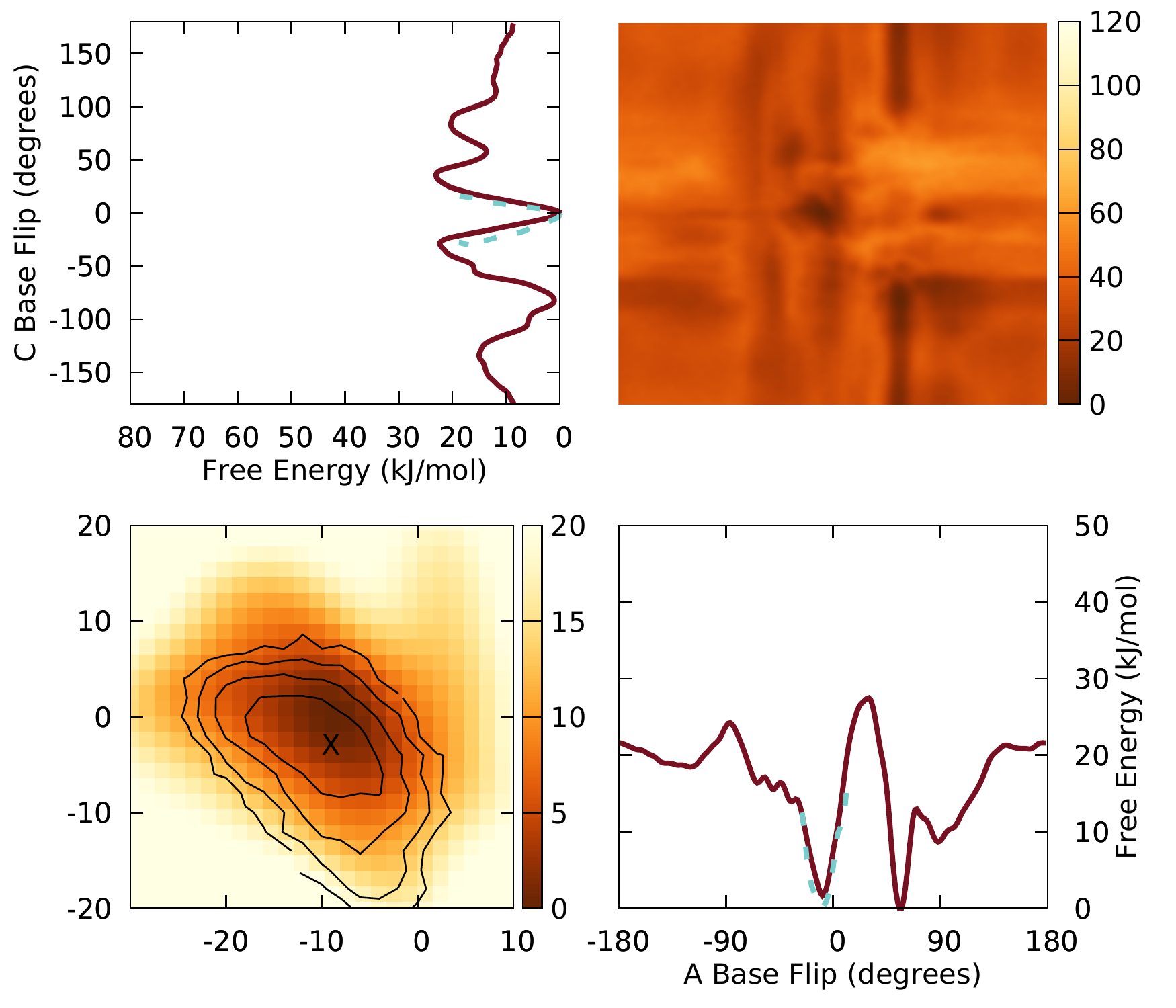}
  \caption{AC}
 \end{subfigure}

 \begin{subfigure}[t]{0.45\textwidth}
  \includegraphics[width=\textwidth]{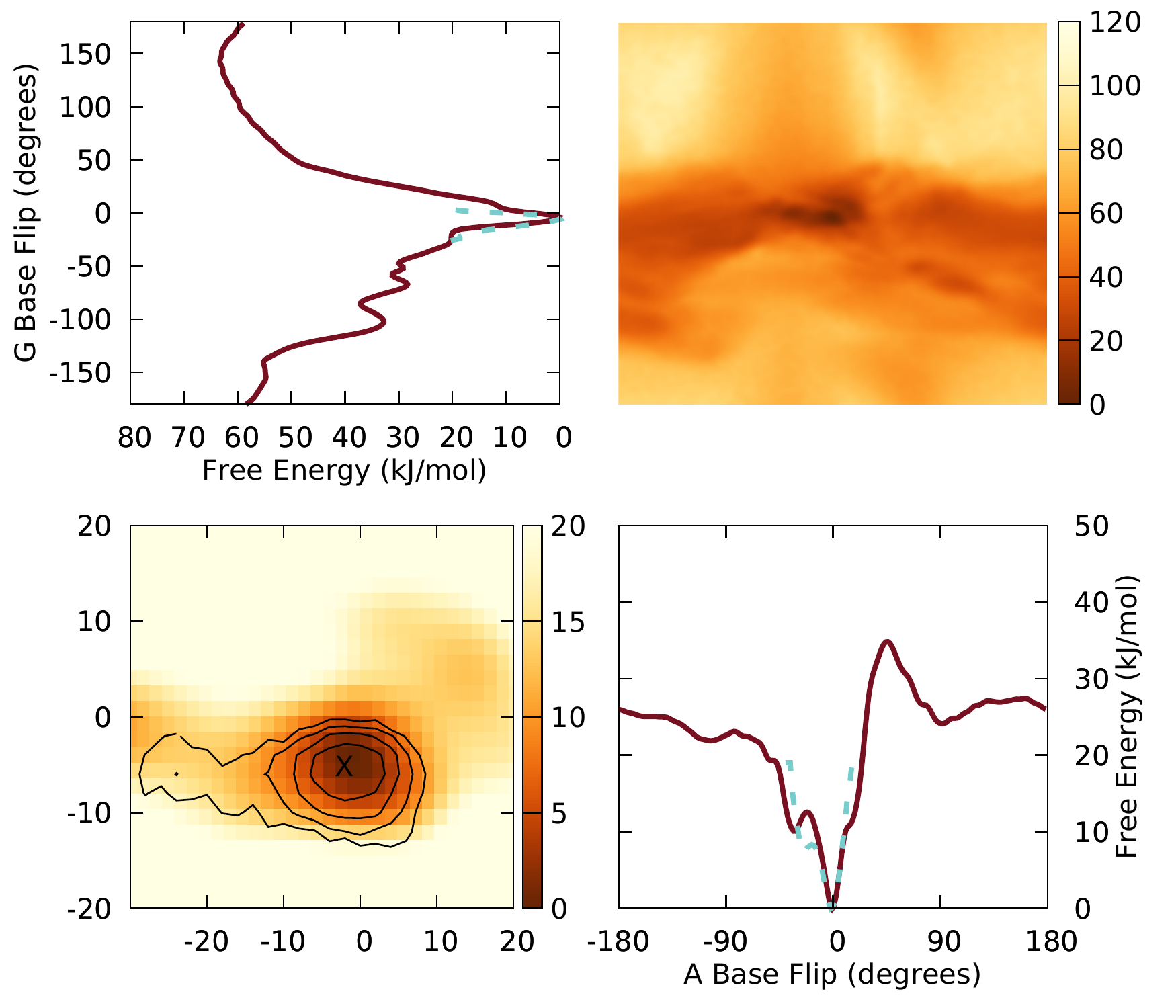}
  \caption{AG}
 \end{subfigure} \quad
 \begin{subfigure}[t]{0.45\textwidth}
  \includegraphics[width=\textwidth]{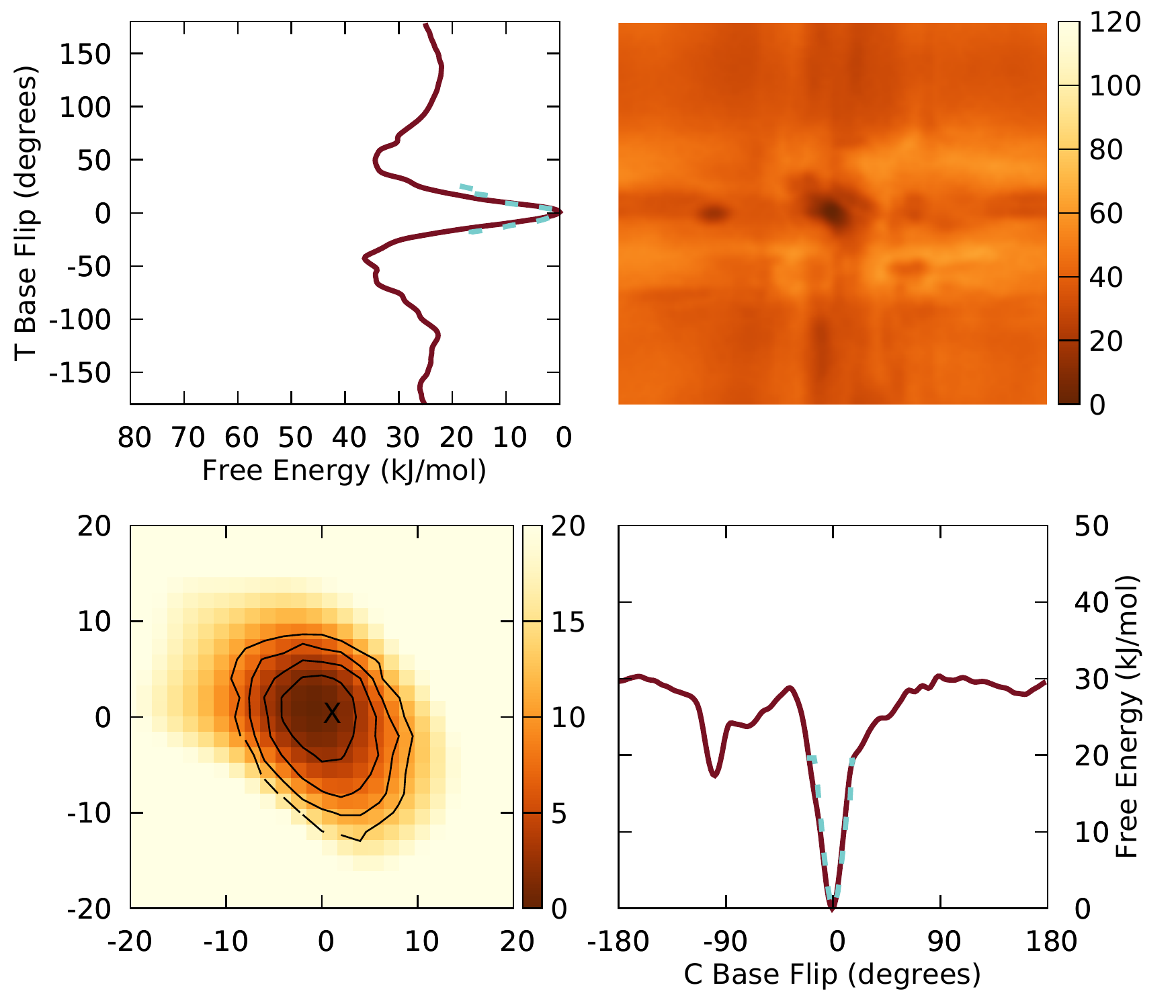}
  \caption{CT}
 \end{subfigure}
 \caption{\label{fig:baseflipAAAGCCCT}
Free energy of base flip for the matched pairs (a) AA, (b) AC, (c) AG, and (d) CT. Figure~\ref{fig:baseflip2D} describes each sub-figure.}
\end{figure}

For most configurations studied, a lower-energy ``plus'' shape is apparent in the 2D FES,
with low-energy states extending both vertically and horizontally from the global minimum.
We display representative free energies in Fig.~\ref{fig:baseflip2D}; the remaining profiles are shown in Figs.~\ref{fig:baseflipTACGGC}~and~\ref{fig:baseflipAAAGCCCT}.
This result indicates it is easier to flip only one base out of the polymer, rather than both simultaneously.
The pairs CC and TT both exhibit two minima at the flipped-in ground states.
We clustered the unbiased trajectories to identify molecular geometries corresponding to these minima:
different hydrogen bonding configurations illustrated in Fig.~\ref{fig:mulmin}.
These ``wobble'' structures have been observed experimentally, not only for TT and CC, but for AA, CT, and GG as well.~\cite{kouchakdjian_pyrimidine_1988,gervais_solution_1995,boulard_solution_1997,peyret_nearest-neighbor_1999,he_preferential_2011}
In contrast, all matched pairs include a single relevant stable state with a nearly-isotropic free energy minimum.
Unlike previous works~\cite{banavali_free_2002, onei_structures_2008, song_improved_2009, imhof_effect_2013, law_base-flipping_2011} which calculated the free energy profiles of one flipping base, we observe a local minimum at roughly (90$^\circ$, 100$^\circ$) in most free energy profiles. This minimum is largely unexpected, especially as it produces a large impact on the difference in free energy, $\Delta G$, between the hybridized and solvent-facing states. (Eq.~\ref{eq:delG})
We display our results in Table~\ref{tab:isoHbond}.
For the AC pair this secondary minimum is so prominent its free energy is lower than for the hybridized state.
We cluster our metadynamics simulations from this minimum and generate a representative image. (Fig.~\ref{fig:secondary_minimum})
Both bases flip out of the DNA polymer and nestle into the minor groove, pointing toward the 5' termini of the DNA.
Because both bases are solvent-facing in this motif, base flip studies biasing only one base could not observe this structure.

This metastable arrangement could be an artifact of the CHARMM27 force field, as it was not parameterized to accurately describe such a structure.
Nevertheless, our arrangement is consistent with the e-motif proposed by Gao et. al.~\cite{gao_new_1995} in their NMR study of mismatched CC pairs; this gives the possibility of our structure being an experimental possibility more weight.
Previous works have investigated the e-motif for mismatch CC pairs in strands with the sequence CCG, typically in triplet repeats i.e. (CCG)$_N$.~\cite{gao_new_1995, zheng_genetically_1996, yu_at_1997, zhang_structure_2017, chou_unusual_2003, romero_anomalous_1999, edfeldt_solution_2004, renciuk_cgg_2011, rojsitthisak_extrahelical_2001}
Several of these works identified intrastrand cross-linking stabilizing the motif,~\cite{chou_unusual_2003, romero_anomalous_1999, edfeldt_solution_2004, rojsitthisak_extrahelical_2001} suggesting our restraint of neighboring base pairs increases the likelihood of observing this motif.
In their molecular dynamics simulation study, Zhang et. al.~\cite{zhang_structure_2017} demonstrate the e-motif is better stacked, more compact, and allows for better packing than a structure not presenting the motif.
We found the e-motif to be stable; a 100~ns unbiased simulation with the e-motif as an initial condition remained in the motif for the entire length of the simulation.
Edfeldt et. al.~\cite{edfeldt_sequence_2004} demonstrated on intrinsic sequence-dependence of the e-motif; we speculate our DNA sequence affects which pairs express or don't express the motif.
CC mismatches are some of the most difficult to be recognized by MMR proteins,~\cite{su_mispair_1988, thomas_heteroduplex_1991} but are the most likely to exhibit the e-motif.~\cite{yu_at_1997, zhang_structure_2017, romero_anomalous_1999}
One might speculate whether there is an inverse relationship between presence of the e-motif and MMR recognition.


\section{Conclusion}

Our results demonstrate large structural differences between matched and mismatched DNA base pairs.
The distributions of order parameters, especially those that measure structure perpendicular to the DNA, differ more for mismatched base pairs than for the corresponding matched pairs.
These parameters include stretch, shear, opening, twist, and shift.
Mismatched base pairs retain fewer inter-base hydrogen bonds than matched ones.
Surprisingly, we observe that structural order parameters
do not seem to follow experimentally-determined MMR recognition trends.~\cite{kramer_different_1984,su_mispair_1988,thomas_heteroduplex_1991,kunkel_eukaryotic_2015}
This is a qualitative result; the studies cited employ different DNA sequences and experimental methodologies, both of which could precipitate significant changes of their observed results.

Enzymes which recognize and repair base pair mismatches flip mismatched bases into their active sites.
For this reason, we rotate bases out of the double helix and into solution as measured by the base flip order parameter.
Our results reveal the base flip angle to be mostly uncorrelated with traditional measures of DNA structure, especially at large flipping angles.

To quantify the work required to flip a base, we compute the free energy profiles of base flip by utilizing the well-tempered metadynamics technique.
Unlike previous works which bias only one base,~\cite{banavali_free_2002, imhof_effect_2013, yin_dynamics_2014} we encounter no systematic difference in the free energy between the hybridized and solvent-facing state for mismatched base pairs than for the corresponding matched pairs. Instead, by analyzing the base flip of both bases simultaneously, we demonstrate the presence of the rare e-motif, in which both bases are solvent-facing, nestled within the minor groove of the DNA.

Multidimensional free energy surfaces reveal the existence of diverse metastable hybridized conformations for some mismatched base pairs.
These states correspond to different hydrogen bonding geometries.
As expected, we do not observe multiple metastable states for the matched Watson-Crick base pairs.

Our metadynamics simulations revealed the presence of the e-motif, but our results illustrate its presence does not correlate directly with experimentally-determined binding affinities of mismatch repair enzymes,~\cite{kramer_different_1984,su_mispair_1988,thomas_heteroduplex_1991,kunkel_eukaryotic_2015} implying this motif is not sufficient to quantitatively determine the efficacy of mismatch detection.
Effects contributing to mismatch detection not incorporated into our study include the direct interaction of DNA and MMR enzymes, as well as enzyme-induced DNA conformations such as bending.~\cite{sharma_differential_2014,sharma_dna_2013,law_base-flipping_2011,vafabakhsh_extreme_2012,erie_single_2014}
The local DNA sequence could greatly alter the free energy landscapes, especially when determining the presence or absence of the e-motif.~\cite{edfeldt_sequence_2004}
Future research will focus on the free energy of flipping mismatched bases in the presence of repair enzymes, and with multiple DNA sequences.


\section{Acknowledgments}

The authors wish to thank Jim Pfaendtner and Patrick Burney for advice on metadynamics simulations.
This work was facilitated though the use of advanced computational, storage, and networking infrastructure provided by the Hyak supercomputer system and funded by the STF at the University of Washington.




\clearpage
\listoffigures


\pagebreak

\clearpage
\setcounter{page}{1}

\pagebreak

\section{DNA Base Pair Mismatches Induce Structural Changes and Alter the Free Energy Landscape of Base Flip: \\ SUPPLEMENTARY INFORMATION}

A. Kingsland$^1$ and L. Maibaum$^1$

$^1$Department of Chemistry, University of Washington, Seattle, WA, USA

\vspace{3em}

\beginsupplement


\section{Supplementary References}

\begin{enumerate}
 \item Lu, X.-J., and W. K. Olson, 2008. 3DNA: A versatile, integrated software system for the analysis, rebuilding and visualization of three-dimensional nucleic-acid structures. \textit{Nat. Protoc.} 3:1213-1227.
\end{enumerate}

\end{document}